\documentclass[%
 reprint,
 superscriptaddress,
 amsmath,amssymb,
prb,
]{revtex4-2}

\usepackage{physics}
\usepackage[utf8]{inputenc}
 
\usepackage{hyperref}
\usepackage{graphicx}
\usepackage{multirow}

\usepackage{amsmath,amssymb,amsfonts,textcomp}

\usepackage{xspace}

\usepackage{siunitx}
\begin{document}

\title{Accurate and Interpretable Representation of  Correlated Electronic Structure via Tensor Product Selected CI}
\author{Nicole M. Braunscheidel}
\affiliation{Department of Chemistry, Virginia Tech,
Blacksburg, VA 24060, USA}
\author{Arnab Bachhar}
\affiliation{Department of Chemistry, Virginia Tech,
Blacksburg, VA 24060, USA}

\author{Nicholas J. Mayhall}
\email{nmayhall@vt.edu}
\affiliation{Department of Chemistry, Virginia Tech,
Blacksburg, VA 24060, USA}

\begin{abstract}
The task of computing wavefunctions that are accurate, yet simple enough mathematical objects to use for reasoning has long been a challenge in quantum chemistry.
The difficulty in drawing physical conclusions from a wavefunction is often related to the generally large number of configurations with similar weights. 
In Tensor Product Selected CI, we use a locally correlated tensor product state basis, which has the effect of concentrating the weight of a state onto a smaller number of physically interpretable degrees of freedom. 
In this paper, we apply TPSCI to a series of three molecular systems ranging in separability, one of which is the first application of TPSCI to an open-shell bimetallic system. 
For each of these systems, we obtain accurate solutions to large active spaces, and analyze the resulting wavefunctions through a series of different approaches including (i)  direct inspection of the TPS basis coefficients, (ii) construction of Bloch effective Hamiltonians, and (iii) computation of cluster correlation functions. 
\end{abstract}

\maketitle
\section{Introduction}

Computational electronic structure theory has developed into an indisputably powerful tool for understanding the quantum mechanical origins of molecular structure and chemical transformations. Progress over the past several decades (in both hardware and algorithmic improvements) has advanced quantum chemistry to the point where the accuracy can often rival that of experiments, particularly for low-energy molecules near equilibrium.  However, as the accuracy of a computation increases, so to does the numerical complexity of the solution, making interpretation more challenging.

The need for achieving both quantitative accuracy and qualitative richness was recognized early on, as computers were first becoming increasingly powerful \cite{ruedenbergAreAtomsIntrinsic1982, ruedenbergAreAtomsIntrinsic1982a, ruedenbergAreAtomsSic1982}.
Gaining access to the underlying driving forces of reactions or structure has proven to be one of the most valuable aspects of quantum chemistry.  As such, the ability to extract qualitative insight is perhaps more important than simply arriving at a quantitatively accurate result. 

Many approaches to extracting conceptual insight from ab initio calculations involves some sort of ``localization''. This is because much of our chemical vocabulary is inherently local (oxidation states, bond order, partial charges, hybridization, etc). The abundance of local chemical constructs is not an accident, molecular structure is generally highly local. For example, the alcohol group in 1-hexanol behaves very similarly to 1-heptanol. As such, understanding the local structure of a functional group in one system extends significant reasoning power to other systems. Consequently, much of the effort spent toward extracting qualitative insight involve the localization of orbitals, such as with NBOs,\cite{fosterNaturalHybridOrbitals1980, glendeningNaturalBondOrbital2012, glendeningNaturalEnergyDecomposition1994, reedNaturalPopulationAnalysis1985} 
ALMOs,\cite{khaliullinEfficientSelfconsistentField2006, khaliullinUnravellingOriginIntermolecular2007, maoIntermolecularInteractionEnergies2021} 
localization of the density as in AIM,\cite{baderQuantumTheoryMolecular1991}
or even many-electron states\cite{caveCalculationElectronicCoupling1997, caveGeneralizationMullikenHushTreatment1996, chenMediatedExcitationEnergy2008, fatehiDerivativeCouplingsAnalytic2013, hsuCharacterizationShortRangeCouplings2008, ruedenbergQuantumChemicalDetermination1993, subotnikConstructingDiabaticStates2008, zhengGeneralizedMullikenHushAnalysis2005}
(though this list is necessarily far from comprehensive).
Localization has also been leveraged extensively for reducing computational complexity. Underlying many of these developments is the fact that the density matrix asymptotically approaches linearly scaling for gapped systems in a localized basis \cite{baerSparsityDensityMatrix1997}. 

All (most) of the methods discussed above ultimately leverage the fact that a Slater determinant wavefunction (or MP2 or CCSD) is invariant with respect to orbital rotations within the occupied or virtual spaces. Orbitals can be mixed to maximize some localizing objective function, and the resulting wavefunctions can then be analyzed in terms of local or non-local contributions. 

In contrast, a tensor product space permits a much more explicit notion of locality, one that naturally exposes the ability to factorize into local quantities (entanglement) and allows clear labeling of the entire Hilbert space in terms of unambiguously local quantities. Recently, we have explored the ability to leverage features of tensor product spaces to decrease computational cost of large active space calculations \cite{abrahamClusterManybodyExpansion2021, abrahamCoupledElectronPairType2022, abrahamSelectedConfigurationInteraction2020c, braunscheidelGeneralizationTensorProduct2023a, mayhallUsingHigherOrderSingular2017a}.
In Refs. \citenum{abrahamSelectedConfigurationInteraction2020c,  braunscheidelGeneralizationTensorProduct2023a}, we introduced a method called Tensor Product Selected Configuration Interaction (TPSCI) which uses a selected CI algorithm to assemble a basis of tensor products of locally correlated wavefunctions, which can provide accurate approximations to FCI. 
In this paper, we demonstrate the ability of these (still expansive) TPSCI wavefunctions to be meaningfully analyzed and interpreted across a rather wide range of physical systems, including non-bonded chromophores, dichromium spin coupling, and the fully delocalized $\pi$ system of a graphene nano-flake. 

\section{Theory}

We will start by expressing the electronic Hamiltonian in a basis of active orbitals ($p,q,r,s$),
\begin{align}
    \hat{H} = h_{pq}\hat{p}^\dagger\hat{q} + \tfrac{1}{2}\left<pq|rs\right>\hat{p}^\dagger\hat{q}^\dagger\hat{s}\hat{r},
\end{align}
assuming that the chosen active space is large enough to capture the necessary physics. \textit{This is generally the most limiting assumption in this paper}, and work to include the dynamical correlation arising from external orbitals is currently underway in our lab \cite{baumanDownfoldingManybodyHamiltonians2019}.

\subsection{Orbital Clustering}
To make progress toward a compact and interpretable representation, we will assume that the active orbitals can be partitioned into disjoint \textbf{clusters}, or groups of orbitals. We will generally use capital letters, $I$, to index clusters. This orbital partitioning (or ``clustering'') is chosen to maximize the interactions within a cluster while minimizing the interactions between clusters. For example, if one had a bimetallic compound (as we consider later in the paper) then one might define each cluster to include all the orbitals centered on a given metal such that all local dynamical correlation is included as intra-cluster correlation, and weak spin-coupling is considered as an inter-cluster correlation.

Physically, we will assume that the interactions within a cluster are stronger than the interactions between clusters. This is not a formal requirement, but rather one that affects the convergence of the calculations.
Each cluster is effectively a new smaller active space, and thus we can construct correlated many-body wavefunctions, $\ket{\alpha_I}$, that are completely localized to each local active space (cluster), $I$.  We will refer to these locally correlated states as \textbf{cluster states}, using them to form an orthonormal basis for the full Fock space on each cluster. 
Likewise, the list of all tensor products of cluster states forms an orthonormal basis for the global Fock space on the full orbital active space. This allows us to represent an arbitrary wavefunction with $s$ states as a linear combination of cluster state tensor products:
\begin{align}
    \ket{\Psi_s} = \sum_{\alpha, \beta, \dots, \gamma}\ket{\alpha_1}\ket{\beta_2}\cdots\ket{\gamma_N}c_{\alpha,\beta,\dots,\gamma}^s.
\end{align}
In this representation, the basis vectors can potentially contain a significant amount of electron correlation folded into the local many-body cluster states. This means that the coefficient tensor, $c^s_{\alpha, \beta, \dots, \gamma}$ only needs to describe \textit{inter}-cluster correlation, with all the \textit{intra}-cluster correlation being folded into the basis vectors. 
This is essentially the same basis used in the Active Space Decomposition (ASD) approach of Shiozaki and coworkers \cite{parkerCommunicationActivespaceDecomposition2013, parkerQuasidiabaticStatesActive2014, parkerModelHamiltonianAnalysis2014}.

\subsubsection{Cluster Mean-Field (cMF) theory}
In order to make the most use out of the representation defined above, it is important that the cluster states are defined carefully, so that they incorporate as much relevant electron correlation as possible. 
Diagonalizing the Hamiltonian projected onto a single cluster (simply keeping only the terms where all creation/annihilation operators act on orbitals within the cluster), yields a set of correlated many-body cluster states that include an exact description of the intra-cluster correlation. Taking a product of the local FCI ground states provides a reasonable approximation for the global ground state, one that becomes exact in the ``clusterable'' limit.

However, interesting molecular systems generally have non-trivial interactions between clusters, and so this becomes a rather poor approximation in practice. Fortunately, one can easily obtain a much improved ground state estimate by including a mean-field description of the inter-cluster interactions when defining the local cluster Hamiltonian, instead of simply projecting out the inter-cluster terms.  An approach, called Cluster Mean-Field (cMF) theory, was introduced by Scuseria and coworkers \cite{jimenez-hoyosClusterbasedMeanfieldPerturbative2015, papastathopoulos-katsarosCoupledClusterPerturbation2022, papastathopoulos-katsarosLinearCombinationsCluster2024, papastathopoulos-katsarosSymmetryprojectedClusterMeanfield2023}, 
and used by Gagliardi and coworkers under the name variational localized active space self-consistent field (vLASSCF) \cite{hermesVariationalLocalizedActive2020}. 
In this work, we construct correlated cluster states by diagonalizing the local cMF effective Hamiltonian, $\hat{H}_I^{cMF}$, for each cluster:
\begin{align}\label{eq:cmf_ham}
    \hat{H}_I^{cMF} 
    = \hat{H}_I + \sum_{J\neq I}\sum_{pq\in I} \sum_{rs\in J}\hat{p}^\dagger\hat{q}(pq||rs)\gamma_{rs}^J,
\end{align}
where $\gamma_{rs}^J = \bra{0_J}\hat{r}^\dagger\hat{s}\ket{0_J}$.
The cMF Hamiltonian for cluster $I$, depends on the 1RDM of all the other clusters, 
requiring the cMF solution to be obtained self-consistently.

Bearing a strong resemblance to traditional Hartree-Fock theory, the self-consistent solution corresponds to the variational minimization of an unentangled (product state)  wavefunction ansatz, 
\begin{align}\label{eqn:cmf}
    \ket{\Psi^{cMF}} =&\ket{0_1}\otimes\ket{0_2}\otimes...\otimes\ket{0_N} \nonumber\\
    =& \ket{0_10_2\dots0_N},
\end{align}
the difference from HF being that in cMF the ansatz only enforces the absence of entanglement between clusters. In further analogy to Hartree-Fock theory, a ``generalized Brillouin condition'' holds, that rigorously uncouples the (converged) cMF wavefunction (Eq. \ref{eqn:cmf}) from tensor product states (TPS's) with a single cluster excited:
\begin{align}\label{eq:brillouin}
    \bra{0_10_2\dots 0_I\dots0_N}\hat{H}\ket{0_10_2\dots\alpha_I\dots 0_N} \nonumber\\
    = \bra{0_I}\hat{H}^{cMF}_I\ket{\alpha_I} = 0.
\end{align}
\subsubsection{Orbital optimization}
Once converged, the cMF energy is stationary with respect to the local cluster state wavefunction coefficients (local FCI coefficients). This means that the cMF energy is invariant to intra-cluster orbital rotations (assuming each cluster is solved exactly), but variant with respect to inter-cluster orbital rotations. In order to obtain a further improved product state wavefunction, we can make the cMF energy stationary with respect to all orbital rotations. This is  essentially a CASSCF calculation with multiple disjoint active spaces \cite{jimenez-hoyosClusterbasedMeanfieldPerturbative2015}. While this clearly has the benefit of providing a lower energy variational solution, perhaps more importantly, is that it removes most of the arbitrariness of the orbital clustering. Assuming each cluster has a Hilbert space dimension greater than one, orbital optimization with a single tensor product state wavefunction will naturally tend to localize the orbitals, so as to maximize electron correlation. Consequently, methods that use the cMF wavefunction as a reference state will be well-defined, not dependent on a particular heuristic for orbital localization \cite{abrahamCoupledElectronPairType2022, papastathopoulos-katsarosCoupledClusterPerturbation2022}.

\subsection{Tensor Product Selected CI (TPSCI)}
While cMF provides a qualitatively attractive approximation for the ground state of a clustered molecular system,  quantitative accuracy is clearly missing due to the neglect of all inter-cluster entanglement.  
Analogous to the common approach of using substituted Slater determinants for a basis, we will use substituted tensor product states as a basis for the full Hilbert space, 
where each TPS is typically taken to be an eigenstate of a cluster's cMF Hamiltonian.
This is similar to a CI analogue of the Block-Correlated Coupled Cluster (BCCC) approach of Li \cite{liBlockcorrelatedCoupledCluster2004}.

The Hartree-Fock based Slater determinant basis and the TPS basis are equivalent, in that both span the full space.
However, as soon as truncations are made, the two bases span different spaces. 
One benefit of working in a TPS basis where the local Hamiltonians are diagonal, is that due to the fact that local correlation is folded into the basis states themselves, the low-energy solutions become more heavily concentrated on a smaller number of basis states. Consequently, computational methods that exploit sparsity (e.g., selected CI) might be expected to be more performant in the correlated TPS basis than in a Slater determinant basis. In Refs. \cite{abrahamSelectedConfigurationInteraction2020c, braunscheidelGeneralizationTensorProduct2023a}
we demonstrated that this is often true, and can sometimes be leveraged for computational benefit. 

\begin{figure}
    \centering
    \includegraphics[width=\linewidth]{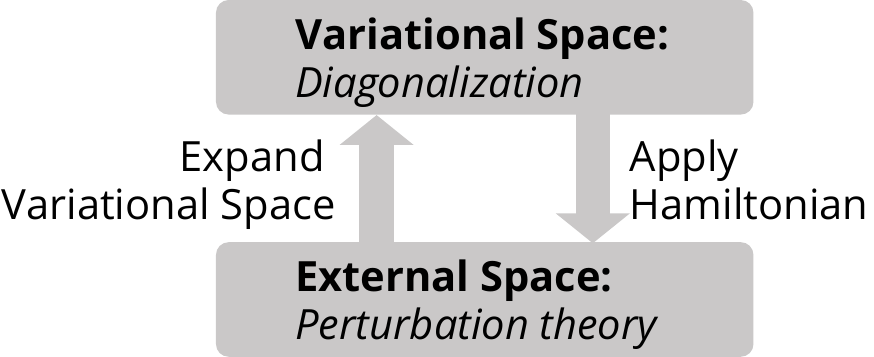}
    \caption{Schematic illustration of the selected CI algorithm used in TPSCI to build a basis of tensor product states. This is iterated until the dimension of the variational space stops growing.}
    \label{fig:cipsi}
\end{figure}

In the Tensor Product Selected CI (TPSCI) method \cite{abrahamSelectedConfigurationInteraction2020c}, we use the general CIPSI \cite{huron1973} algorithm to discover and exploit, in a bottom-up fashion, the sparsity of the exact wavefunction in a TPS basis. 
This uses perturbation theory to iteratively discover the non-negligible TPS's that are needed to accurately approximate the exact solution. This is done by the following steps:
\begin{enumerate}
    \item Diagonalize $\hat{H}$ in the current variational space (this being a list of TPS basis states that are expected to have large amplitude in the exact solution). 
    \item Apply the $\hat{H}$ to the current variational space eigenvector. This couples the variational space to the external space. 
    \item Compute the first order wavefunction in the external space. 
    \item Move the external configurations with large first-order coefficients from the external space to the variational space. 
    \item  If the variational dimension increases, go back to step 1. If not, exit.   
\end{enumerate}
This overall iterative loop is shown in Fig. \ref{fig:cipsi}.
In principle, all local FCI cluster states would be computed and used to form the basis for state space. While this is tractable for small clusters, for larger clusters, this becomes computationally prohibitive, and high-energy states are generally discarded prior to running TPSCI. We generally use $M$ to refer to the maximum number of cluster states kept in a particular fock sector of a cluster. For each particle number subspace included, the corresponding lowest $M$ eigenstates are computed. Then the $\hat{S}^+$ and $\hat{S}^-$ operators are applied to those cluster states to generate the basis for the higher $M_s$ sectors. More details can be found in Ref. \cite{braunscheidelGeneralizationTensorProduct2023a}.

The computational limitations of conventional (determinant basis) CIPSI  is generally determined  by the size of the dimension of the variational space. Because TPSCI uses correlated TPS basis vectors, more correlation energy is typically recovered with smaller variational spaces, addressing the most significant bottleneck. This comes at a cost, however, during the matrix element evaluation. Whereas evaluating matrix elements in the Slater determinant basis is extremely efficient, matrix elements in the TPS basis are significantly more expensive. For a 3 cluster example, consider  the following Hamiltonian contribution that contains three operators ($\hat{p}^\dagger\hat{q}^\dagger r$) on cluster 1, and one operator ($s$) on cluster 2:
\begin{align}
    \hat{H}\leftarrow \sum_{pqr}^{\in 1} \sum_{s}^{ \in 2} \left<pq|rs\right> \hat{p}^\dagger\hat{q}^\dagger\hat{s}\hat{r}.
\end{align}
Computing the matrix element of this particular Hamiltonian contribution between two arbitrary TPS basis vectors will require the contraction of an integral sub-block with tensors of local quantities:
\begin{align}
\bra{\alpha'_1\beta'_2\gamma_3} \hat{H}_{1,2} \ket{\alpha_1\beta_2\gamma_3} 
\leftarrow \chi \sum_{pqr}^{\in 1} \sum_{s}^{ \in 2} \left<pq|rs\right> {}^1\Gamma_{pqr}^{\alpha'\alpha} {}^2\Gamma_{s}^{\beta'\beta}
\end{align}
where $\chi$ is a sign determined by the number of electrons in state $\ket{\alpha_1}$, and the $\Gamma$ tensors are the precomputed local operator matrices in the cluster basis, e.g.:
\begin{align}
    {}^1\Gamma_{pqr}^{\alpha'\alpha} = \bra{\alpha'_1}\hat{p}^\dagger\hat{q}^\dagger\hat{r}\ket{\alpha_1}.
\end{align}
Because the matrix elements require a series of tensor contractions, instead of just a single access from an array, the construction of matrix elements becomes the key bottleneck in TPSCI. However, in Ref. \cite{abrahamSelectedConfigurationInteraction2020c}, we compared TPSCI to Heat Bath CI \cite{holmesHeatBathConfigurationInteraction2016}, and found that we were able to obtain significantly lower variational energies using TPSCI than with heat bath CI. 
Once the TPSCI variational space has converged, we have also found that it is sometimes beneficial (especially for ground state problems) to perform a higher-order singular value decomposition (HOSVD) of the resulting wavefunction tensor, to rotate the cluster states into a form that diagonalizes the local cluster reduced density matrices (within subspaces that preserve local particle number and $\hat{S}^z$). 
More details about the implementation and matrix element construction can be found in Refs. \cite{abrahamSelectedConfigurationInteraction2020c, braunscheidelGeneralizationTensorProduct2023a}.

\subsubsection{Bloch Effective Hamiltonian\label{sec:bloch}}
Once an accurate TPSCI wavefunction is obtained, one is often interested in more than just the associated energy. Being able to extract qualitative information to aid in communicating and reasoning about the underlying electronic structure, is extremely valuable, and TPS wavefunctions are uniquely interpretable.
Since the basis states in TPSCI are essentially diabatic states, it is a very natural extension to use the concept of Bloch effective Hamiltonians to extract quantitative relationships between qualitatively meaningful degrees of freedom
\cite{blochTheoriePerturbationsEtats1958, 
descloizeauxSpinWaveSpectrumAntiferromagnetic1962,
calzadoAnalysisMagneticCoupling2002, 
mauriceUniversalTheoreticalApproach2009, 
monariDeterminationSpinHamiltonians2010, 
malrieuMagneticInteractionsMolecules2014,
mayhallSpinflipNonorthogonalConfiguration2014a,
mayhallComputationalQuantumChemistry2015a, 
mayhallModelHamiltoniansInitio2016a, 
pokhilkoEffectiveHamiltoniansDerived2020} for analysis.

We start by defining a ``model space'', $\{\ket{\phi_i}\}$, which is taken to be the set of physically meaningful TPS's that qualitatively define the structure or process. 
 To ensure that the model space is actually relevant to the physics computed, the exact low-energy states of the system, $\ket{\Psi_s}$, should have relatively large projections onto the model space,
i.e., 
\begin{align}\label{eq:norm}
    \norm{\hat{P}_\mathcal{M}\ket{\Psi_s}}\approx 1,
\end{align}
where, $\hat{P}_\mathcal{M}=\sum_i\dyad{\phi_i}{\phi_i}$.
Next, we seek a hermitian effective Hamiltonian which exists only in the model space, but that yields the exact energy spectrum. While this is often done in a bottom-up fashion through approaches like quasi-degenerate perturbation theories or Schrieffer–Wolff transformations, if  one already has access to the exact target eigenstates, an effective Hamiltonian (specifically, a Bloch effective Hamiltonian) can be obtained simply in a top-down fashion by direct projection,
\begin{align}
\hat{H}^{Bloch} =& \ket{\tilde{\Psi}_s}E_s\bra{\tilde{\Psi}_s}\\
=& \ket{\hat{P}_\mathcal{M}\Psi_s}X_{st}E_tX_{tu}\bra{\hat{P}_\mathcal{M}{\Psi}_u},
\end{align}
where, 
\begin{align}
    X_{st} = \left(\bra{\Psi_s}\hat{P}_\mathcal{M}\ket{\Psi_t}\right)^{-1/2},
\end{align}
will always exist when Eq. \ref{eq:norm} holds.
The individual matrix elements of $\hat{H}^{Bloch}$ then contain quantitative  relationships between qualitatively meaningful states. 

\subsubsection{Cluster correlation functions}\label{sec:cumulants}
In addition to the Bloch effective Hamiltonian, which gives us a state-universal description of the interactions between physically intuitive degrees of freedom, we also often want to characterize specific states in terms of physically intuitive variables. 

Following the  recent work of Luzanov, Krylov, and Casanova \cite{casanovaQuantifyingLocalExciton2016a, luzanovQuantifyingChargeResonance2015a},
the local TPS representation makes it simple to compute cluster correlation functions of various local cluster operators to characterize states in terms of observables. A two-cluster correlation function for operator $\hat{O}$ is given as the covariance between the operator localized onto each individual cluster:
\begin{align}
 \textrm{cov}\left(\hat{O}_I,\hat{O}_J\right) =& \bra{\Psi_s}\hat{O}_I\hat{O}_J\ket{\Psi_s}\\ &- \bra{\Psi_s}\hat{O}_I\ket{\Psi_s}\bra{\Psi_s}\hat{O}_J\ket{\Psi_s},
\end{align}
where the covariance of an operator with itself is just the variance, which will be used to measure the local fluctuations in a given cluster. 
Depending on the system we will consider correlations between the following cluster operators: local charge (particle number), $\hat{N}_I$, local spin projection, $\hat{S}^z_I$, local spin $\hat{S}^2_I$
, and local excitation $\hat{Q}_I=\hat{1}-\dyad{0_I}$ which indicates that cluster $I$ is excited out of its cMF ground cluster state. 

\section{Results}
In the following sections, we explore the ability to simultaneously obtain quantitative yet interpretable approximations to large active spaces. 
Because our representation is ideal for separable clusters, to obtain insight into the transferability of the formalism, we explore systems which span a broad spectrum in terms of clusterability, going from  a non-bonded tetracene tetramer (Sec. \ref{sec:sf}),
to an anti-ferromagnetically coupled dichromium complex (Sec. \ref{sec:cr2}), to a completely delocalized graphene-flake model, hexabenzocoronene (Sec. \ref{sec:hbc}).

We use PySCF software for performing any necessary geometry optimizations, Hartree-Fock calculations, and integral generation \cite{sun_recent_2020}. All TPSCI calculations are performed using our open-source FermiCG software \cite{mayhall_nicholas_fermicg_nodate}.

\subsection{Singlet-Fission: Tetracene Tetramer}\label{sec:sf}
\begin{figure*}
    \centering
    \includegraphics[width=\linewidth]{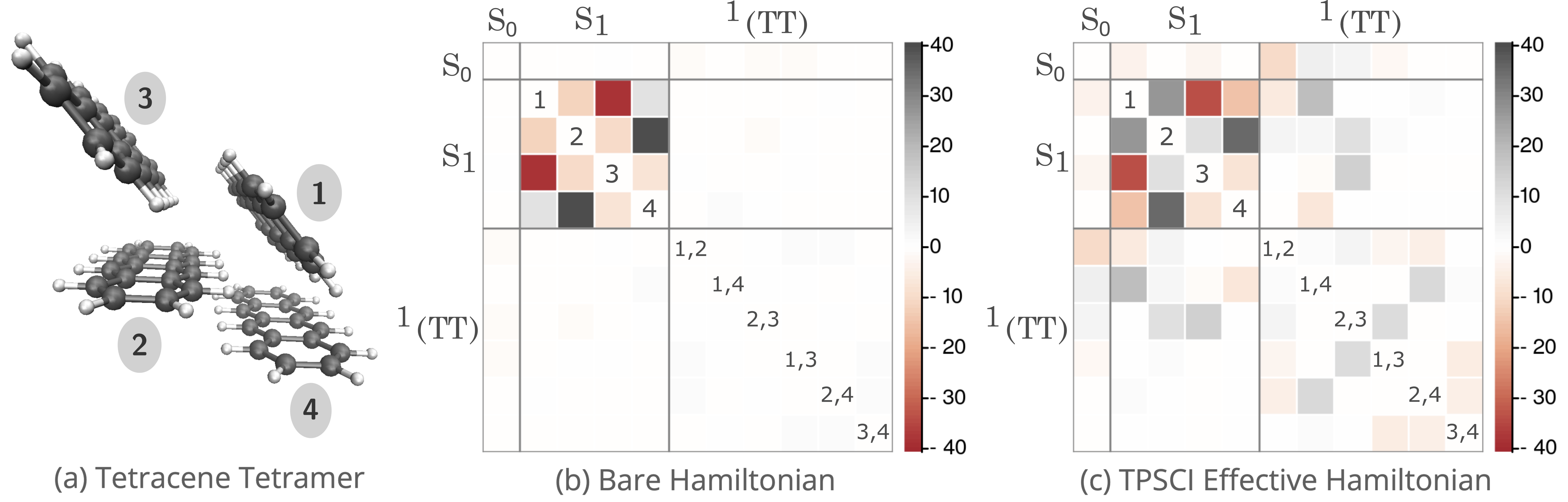}
    \caption{Bare and Bloch Effective Hamiltonians for tetracene tetramer with diagonal entries subtracted to show off-diagonal couplings in meV. Active space of (40e, 40o) with system and clusters labeled. (a) Tetracene tetramer with the associated clusters labeled (b) Hamiltonian matrix for the model space (diabatic basis)  (c) Bloch Effective Hamiltonian obtained by projecting TPSCI wavefunctions onto the model space. The non-singlet states have been removed for clarity.}
    \label{fig:heff_tt}
\end{figure*}

Singlet fission is the photophysical process by which a bright singly excited state, $\ket{\mathrm{S}_1}$, is converted into two lower energy triplet states, $\ket{\mathrm{T}_1} + \ket{\mathrm{T}_1}$, by way of a multiexcitonic intermediate state  $\ket{^1\mathrm{TT}}$ ~\cite{Smith2013}.
Because this process converts a single photon into two excitons (each of which can split into charge carriers), materials that exhibit singlet fission have promising applications in solar cells due to the possibility of overcoming the Shockley--Queisser limit for efficiency\cite{Hanna2006}.

A dimer model description of the singlet-fission process  requires a total of 8 states:
the ground state $\ket{\mathrm{S}_0\mathrm{S}_0}$, two local bright states, $\ket{\mathrm{S}_1\mathrm{S}_0}$ and $\ket{\mathrm{S}_0\mathrm{S}_1}$, two local triplet states, $\ket{\mathrm{S}_0\mathrm{T}_1}$ and$ \ket{\mathrm{T}_1\mathrm{S}_0}$, and the three biexcitonic states arising from the product of two triplet states  $\ket{\mathrm{T}_1\mathrm{T}_1}$.  
The biexcitonic state is characterized as an entangled pair of local triplets, which can be spin-coupled into either a singlet $\ket{^1\mathrm{TT}}$, triplet $\ket{^3\mathrm{TT}}$, or quintet $\ket{^5\mathrm{TT}}$, 
represented in the diabatic basis via their 
Clebsch--Gordan coefficients, 
\begin{align}
    \ket{^1\mathrm{TT}} &= \frac{1}{\sqrt{3}}(\ket{\mathrm{T}^+\mathrm{T}^-} - \ket{\mathrm{T}^0\mathrm{T}^0} + \ket{\mathrm{T}^-\mathrm{T}^+})\\
    \ket{^3\mathrm{TT}} &= \frac{1}{\sqrt{2}}(\ket{\mathrm{T}^+\mathrm{T}^-} - \ket{\mathrm{T}^-\mathrm{T}^+})\\
    \ket{^5\mathrm{TT}} &= \frac{1}{\sqrt{6}}(\ket{\mathrm{T}^+\mathrm{T}^-} - 2\ket{\mathrm{T}^0\mathrm{T}^0} + \ket{\mathrm{T}^-\mathrm{T}^+})
\end{align}
where the first triplet in each state refers to chromophore A and second triplet to chromophore B. 
Although $\ket{^1\mathrm{TT}}$ is the main intermediate, as it is spin-allowed, it has been shown that the triplet and quintet states can play a role in the separation process\cite{johnsonOpenQuestionsPhotophysics2021}.

While a dimer model captures the key intermediates, it is too small to describe additional physical effects that occur in bulk systems. For example, the initial bright state is generally understood to delocalize over several monomers, 
increasing the number of  localized biexcitons to which it can couple.
Further, it has been seen that singlet fission rates are increased by the involvement of a non-nearest neighbor biexciton $\ket{^1\mathrm{T}\cdots \mathrm{T}}$, which is absent from the dimer model by construction
\cite{pensackExcitonDelocalizationDrives2015, scholesCorrelatedPairStates2015}.
Recent work has further demonstrated the importance of beyond dimer effects \cite{berkelbach2014, berkelbach2017}.

Because methods that rely on single or even double excitations struggle to accurately describe two-electron excitations, multireference methods, such as CASPT2, are often required to capture the wide range of electronic character. 
While this is suitable for a couple chomophores, active space methods typically grow exponentially with the number of chromophores, making it difficult to extend to larger systems.

TPSCI (similar to ASD which preceded it \cite{parkerCommunicationActivespaceDecomposition2013,parkerModelHamiltonianAnalysis2014, parkerQuasidiabaticStatesActive2014}) is well suited for treating collections of chromophores because the physical system efficiently maps onto the tensor product basis. 
In a recent paper \cite{braunscheidelGeneralizationTensorProduct2023a}, we demonstrated that TPSCI was able to provide accurate approximations to  CASCI for a large (40e, 40o) active space, which incorporated 10 active orbitals for each of the four tetracene chromophores. 
In this paper, we explore this further, using the TPS structure to facilitate further analysis and characterization of the resulting wavefunctions. 

In this subsection, we use TPSCI to go beyond the minimal dimer example for singlet fission and explore dressed Hamiltonains and local cluster operator correlations for tetracene tetramer (shown in Fig. \ref{fig:heff_tt}(a)) using a large (40e, 40o) active space.

\subsubsection{Active Space Selection and Clustering}
In order to construct an active space which contains the relevant orbitals for describing both the local $S_1$ and $T_1$ states, we used the CIS-NO~\cite{shuConfigurationInteractionSingles2015} approach to select our active space in the 6-31G*\cite{ditchfield1971a} basis. We first performed a CIS calculation for the first singlet and triplet on each chromophore and averaged these states into the one particle reduced density matrix (1RDM). By diagonalizing the 1RDM, we obtained a set of natural orbitals from which we extracted the 40 most correlated orbitals (i.e. those that have the most fractional occupancy) as our active space. We then localized these 40 orbitals using Pipek-Mezey~\cite{PM} and then grouped the orbitals into four (10e, 10o) clusters on each chromophore for an overall active space of 40 orbitals and 40 electrons (40e, 40o). 
We are currently developing approximate solvers (such as RAS-CI) for obtaining the local cluster states. This will allow us to consider clusters larger than the relatively small ten orbital clusters used here.

\subsubsection{Bloch Effective Hamiltonian}
To analyze our TPSCI results, we start by computing a Bloch effective Hamiltonian by projecting the TPSCI eigenvectors onto our diabatic basis (i.e. model space).
The diabatic basis for tetracene tetramer includes the biexciton diabatic states for each pair in addition to the singly excited states on each chromophore.
In the tetramer, there are six possible dimer configurations and each generates three $M_s=0$ spin components $\ket{\mathrm{T}^+\mathrm{T}^-}$, $\ket{\mathrm{T}^0\mathrm{T}^0}$, and $\ket{\mathrm{T}^-\mathrm{T}^+}$ which leads to a total of 18 diabatic biexciton states. 
We also observe in the tetracene monomer that both $\mathrm{T}_1$ and $\mathrm{T}_2$ are lower in energy than the first singlet excited state $\mathrm{S}_1$, thus all three of these states must be represented for each chromophore in our model space.
In total, our model space includes 31 diabatic states.
However, to simplify the picture, we focus on the singlet model space, where the biexcitonic states have been mixed using their Clebsch-Gordon coefficients to form proper $\ket{^1\mathrm{TT}}$ diabatic states.
This reduces our model space from 31 states to 11 states.

In Fig. \ref{fig:heff_tt}(b) and \ref{fig:heff_tt}(c), we plot the bare Hamiltonian and effective Hamiltonian in the model space as described above, with the columns arranged to correspond with the cluster labels in \ref{fig:heff_tt}(a). 
The diagonal energies are subtracted to better reveal the off diagonal activity in meV. 
In both plots, the Hamiltonian is blocked by state type: ground state, four singlet excited states, and the 6 singlet biexcitons, where the non-singlet states have been omitted for clarity.
As expected the singlet excitons couple very strongly to each other, both in the bare and effective Hamiltonians, which ultimately gives rise to bright state delocalization.
There is only negligible coupling between the $S_1$ and biexcitons in the bare plot, but after including external space correlations, we see a significant growth in the strength of the effective coupling.
These are listed explicitly in Table \ref{tbl:s1_tt}.
Whereas the $\ket{^1\mathrm{TT}}$ states on the 3 herringbone dimers develop significant $\mathrm{S}_1$ coupling after the inclusion of the external space, the planar dimers remain uncoupled from the bright spectrum. 
In addition to strengthening the coupling between the bright states and the biexcitonic states, the inclusion of higher energy states also induces couplings between the biexcitons themselves.

\begin{table}[]
\begin{center}
\caption{TPSCI Effective Hamiltonian to show coupling strengths between $\mathrm{S}_1$ and ${}^1(\mathrm{TT})$ in meV.}
\label{tbl:s1_tt}
\begin{tabular}{|c|c|cccccc|}
\hline
  $\hat{H}^{eff}$&&\multicolumn{6}{c|}{$\ket{{}^1\mathrm{TT}}$} \\\hline 		
	&		&	1,2	&	1,4	&	2,3	&	1,3	&	2,4	&	3,4	\\ \hline
 \multirow{5}{*}{$\bra{\textrm{S}_1}$}
	&	1	&	-6.43	&	19.46	&	0.33	&	0.80	&	1.31	&	0.01	\\
	&	2	&	3.66	&	2.47	&	10.16	&	1.04	&	0.50	&	0.00	\\
	&	3	&	0.37	&	-0.82	&	14.36	&	-0.64	&	-0.08	&	-0.07	\\
	&   4	&	-0.23	&	-6.62	&	0.42	&	0.10	&	0.37	&	0.08	\\ \cline{1-8}
\end{tabular}
\end{center}
\end{table}


\begin{figure*}
    \includegraphics[width=\linewidth]{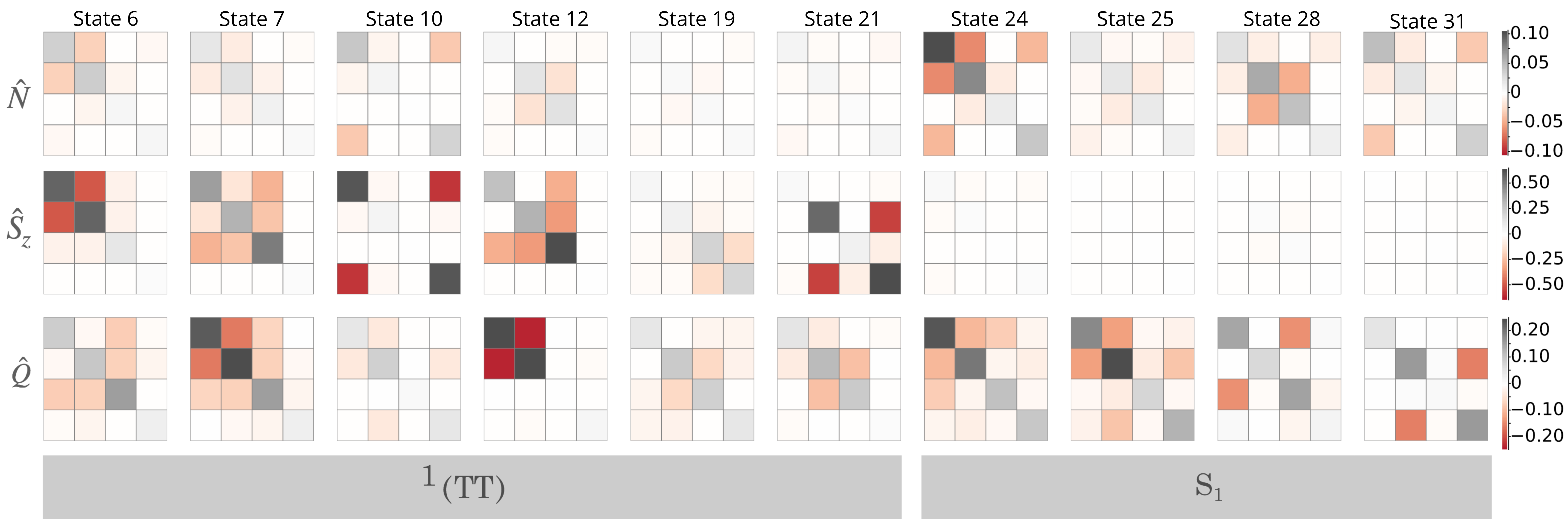}
    \caption{Tetracene tetramer correlation functions. (top row) particle number, $\hat{N}_I$. (middle row) spin projection, $\hat{S}^z_I$. (bottom row) cluster excitation, $\hat{Q}_I$. 
    The first 6 plots from the left are for the $\ket{^1\mathrm{TT}}$ states. 
    The last 4 plots from the left are for the $\ket{\mathrm{S}_1}$ states.
    Each matrix column/row corresponds to cluster 1 to 4, as labeled in Fig. \ref{fig:heff_tt}. 
    The color scale for each correlation function is shown on the far right.
    }
    \label{fig:s1_tt}
\end{figure*}

\begin{table}
\centering
\caption{Tetracene tetramer local excitation strengths.
Columns correspond to the expectation values of the local cluster excitation operator,
$\hat{Q}_I=\hat{1}-\dyad{0_I}.$ 
A value of 0 indicates the cluster is in its local ground state.
A value of 1 indicates that the cluster is always in a local excited state.
Summing all the local excitation rank of the state.
}
\label{tbl:sf_q1}
\begin{tabular}{|l|cccc|c|}  
\hline\hline
 State & $\left<\hat{Q}_1\right>$& $\left<\hat{Q}_2\right>$& $\left<\hat{Q}_3\right>$& 
 $\left<\hat{Q}_4\right>$ & Excitation rank\\ \hline
6 & 0.90 & 0.89 & 0.20 & 0.04 & 2.01 \\
7 & 0.63 & 0.54 & 0.81 & 0.03 & 2.02 \\
10 & 0.95 & 0.10 & 0.01 & 0.95 & 2.01 \\
12 & 0.45 & 0.54 & 1.00 & 0.02 & 2.01 \\
19 & 0.05 & 0.11 & 0.91 & 0.95 & 2.02 \\
21 & 0.05 & 0.86 & 0.11 & 0.99 & 2.01 \\
\hline
24 & 0.63 & 0.27 & 0.13 & 0.11 & 1.14 \\
25 & 0.24 & 0.57 & 0.08 & 0.16 & 1.05 \\
28 & 0.18 & 0.08 & 0.81 & 0.02 & 1.09 \\
31 & 0.06 & 0.20 & 0.01 & 0.80 & 1.07 \\
\hline\hline
\end{tabular}
\end{table}

\subsubsection{Correlation Analysis}
As mentioned in Section\ref{sec:cumulants}, correlation functions of various local cluster operators can be used to characterize  the adiabatic TPSCI wavefunction in terms of physically meaningful  relationships between clusters. 
In Table \ref{tbl:sf_q1}, we list the expectation values of the cluster excitation operators, which measures the amount of excited state character in each state, and by summing over clusters, the total excitation rank of each excited state.
This allows us to unambiguously identify the biexcitons, which we use to label the states accordingly in Fig. \ref{fig:s1_tt}, where we compute the inter-cluster cumulants for  cluster particle numbers ($\hat{N}_I$), cluster spin projections ($\hat{S}^z_I$), and cluster excitations ($\hat{Q}_I$)
for each of the six singlet biexcitons ${}^1(\mathrm{TT})$ and four bright states.


Looking first at the particle number correlations, $\hat{N}_I$, we see that overall, the ``dark'' $\ket{\mathrm{TT}}$ states are relatively quiet compared to the charge correlations present in the bright states.
This is entirely expected based on the physical characteristics of the bright vs dark states. Bright states have relatively large amounts of charge transfer character mixed in.
The presence of charge transfer makes local particle number less well defined, which increases a cluster's charge variance, and similarly increases the anti-correlation between two clusters' charge states (when the donor is cationic, the accepter has a high probability of being anionic). 
Although weaker than the lowest $\mathrm{S}_1$ state, we see clear signatures of charge correlation  present in a couple of the biexcitons (State 6 and 10).
This is consistent with an analysis of the wavefunction itself.
If we compute the amount of charge transfer present in each state, we find that out of all of the biexcitons, states 6 and 10 have the highest percentage of CT character, 4.6\% and 4.5\%, respectively (see the Supplementary Information for all state CT compositions). 
Charge correlations between clusters entangled into a biexciton also indicates significant superexchange, which stabilizes the low spin biexciton \cite{berkelbachMicroscopicTheorySinglet2013b,abrahamSimpleRulePredict2017a}.

Considering next the $\hat{S}^z_I$ correlations, we see the opposite trend, where the biexciton states have significantly more pronounced correlations, and the bright states are featureless (as would be expected from the lack of local triplet character). 
This is also consistent with the nature of the different sets of states. 
A $\ket{^1\mathrm{TT}}$ biexciton is characterized as two entangled triplet states coupled into a singlet state. Because the total $M_s$ is zero, when the first monomer is in the $M_s=1$ microstate, the entangled partner is very likely to be found in the $M_s=-1$ microstate. This entanglement leads to a very strong $\hat{S}^z_I$ covariance. 

Using the spin correlation  as a way to label the biexcitons \cite{casanovaQuantifyingLocalExciton2016a}, 
we can identify that  state 6 is a (1,2) biexciton (meaning that it primarily exists on chromophores 1 and 2), states 7 and 12 are superpositions of (1,3) and (2,3) biexcitons, and state 21 is a (2,4) biexciton. 
Looking at state 19, we see what resembles a (3,4) biexciton, although the overall magnitude is much smaller. 
In order to understand this, we can look at the $\hat{S}^2$ expectation value of the state (shown in the Supplementary Information). We find that in this case, the (what we labeled to be ) $\ket{^1\mathrm{TT}}$ state has significant spin contamination of about 1.1. This is a consequence of the fact that the , $\ket{^1\mathrm{TT}}$ and $\ket{^3\mathrm{TT}}$ states are approximately degenerate, meaning that any arbitrary mixing of the two states is also an eigenstate. This mixing of the two spin states essentially creates a ``broken-symmetry'' state, where one chromophore is $M_s=1$ and the other is $M_s=-1$. By locking the local spin vectors, the local $\hat{S}^z_I$ fluctuations are diminished, and hence the ability to have significant covariance with any other cluster. This could be corrected by tightening our TPSCI convergence, or by rediagonalizing $\hat{S}^2$ in nearly-degenerate subspaces. 

Looking more closely at state 12, we see that clusters 1, 2, and 3 all have significant $\hat{S}^z_I$ fluctuations, but only the 1,3 and 2,3 pairs are spin correlated.
Considering the cluster excitation, $\hat{Q}_I$ covariance plot, we see that while cluster three has zero fluctuations (it is consistently excited), clusters 1 and 2 are strongly correlated. This means that cluster 3 is always excited, but when cluster 1 is excited, cluster 2 is most likely in the ground state, and vice versa. 
This suggests a situation where a triplet state on cluster 3 forms a biexciton with a triplet delocalized between clusters 1 and 2.  
This is consistent with an analysis of the individual TPS state coefficients, were we find that 98\% of the wavefunction is characterized as a superposition of the 1,3 and 2,3 biexcitons,  $\ket{\Psi_{12}}\approx 0.67\ket{^1(\mathrm{T}_1\mathrm{S}_0\mathrm{T}_1\mathrm{S}_0)}+0.73\ket{^1(\mathrm{S}_0\mathrm{T}_1\mathrm{T}_1\mathrm{S}_0)}$.

\subsection{Cr$_2$ Complex Effective Hamiltonian}\label{sec:cr2}
Multi-center organometallic complexes often exhibit interesting physics, such as single molecule magnetism\cite{caneschiAlternatingCurrentSusceptibility1991, demirGiantCoercivityHigh2017}, or valuable catalytic capabilities, such as water oxidation\cite{Gil-Sepulcre2022}, or nitrogen fixation\cite{TANABE2022214783}.
While having a computational method that could efficiently compute the low-energy structure of organometallic compounds would be highly valuable, several physical features of these systems make this challenging.  When multiple weakly interacting metal centers possess unpaired electrons, the resulting low-energy states are highly multiconfigurational, making conventional approaches like perturbation theory
or coupled cluster theory inappropriate, as they require a qualitatively correct single determinant wavefunction as a reference. 

\begin{figure}
    \centering
    \includegraphics[width=.6\linewidth]{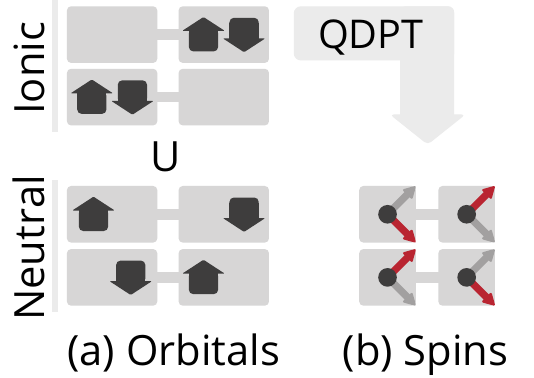}
    \caption{Cartoon illustration of the electronic structure of an $S=\tfrac{1}{2}$ biradical approximately mapped onto an isotropic Heisenberg model at the large $U$ limit, both restricted to the relevant $M_s=0$ subspace. (a) the 4 possible Slater determinants. Assuming a localized orbital basis, the bottom two correspond to neutral excitations and the top two are ionic or charge transfer configurations. (b): the 2 possible spin configurations after being mapped onto a Heisenberg model with quasi-degenerate perturbation theory (QDPT).}
    \label{fig:biradical}
\end{figure}

Because the product of multiple high-spin centers leads to a large number of spin states, 
it is not always possible to predict, a priori, the spin multiplicity of a multi-center organometallic complex. 
Consider a simple biradical Hubbard model, represented in a local minimal orbital basis, as shown in Fig. \ref{fig:biradical}(a). Here, the low-energy configurations are both open-shell broken symmetry states, and proper eigenstates must be superpositions of these two configurations, one being a singlet, and the other the $M_s=0$ component of the triplet. Because the ionic (or charge transfer) configurations are both singlets, they can only mix with the singlet combination of neutral configurations, which is ultimately the origin of the antiferromagnetic coupling in biradicals. 

This picture can often be simplified significantly. For systems where the electron repulsion is sufficiently large such that hopping of an electron from one center to another incurs a high energy barrier (i.e., $U\gg t$),  the influence of charge fluctuations can be approximately downfolded using quasi-degenerate perturbation theory into an effective spin Hamiltonian, called the Heisenberg-Dirac-van Vleck Hamiltonian:\footnote{This is true for the Hubbard model at half-filling. Situations with away-from half-filling can be mapped onto different Hamiltonians, such as the $tJ$ model.}
\begin{align}
    \hat{H}^{HDvV} = -2J\hat{S}_1\cdot\hat{S}_2,
\end{align}
where $J=-\tfrac{4t^2}{U}$.
For this model,  the low-energy spectrum is completely determined by the value of $J$. 

Starting from the ab initio Hamiltonian instead of the Hubbard model, one finds that the zeroth-order term also contains the non-local direct exchange integral. Since bare exchange stabilizes the high-spin states, and the second order super-exchange term stabilizes the low-spin states, even getting  the sign correct for the exchange coupling constant can be difficult, as the value of $J$ is determined by the subtle interactions coupling the metals with each other and with the ligands. 
However, once known, the relative ordering of the spin states can be directly written down in terms of $J$. While this is indeed an approximate description of the low-energy electronic structure, it is profoundly useful as the prediction of $J$ is also often what connects experiment to theory, where $J$ is commonly fit to experimental magnetic susceptibility measurements.  

The most common approach to computing $J$ from ab initio quantum chemistry is to use DFT, where one of the degenerate broken symmetry configurations is optimized, followed by a spin projection formula, originally proposed by Noodleman \cite{noodlemanValenceBondDescription1981}, and then improved by Yamaguchi \cite{yamaguchiSpinCorrectionProcedure1988, yamanakaEffectiveExchangeIntegrals1994a}.  
While this approach has been widely used due to its conceptual simplicity and computational efficiency, there are downsides. First, the formalism doesn't actually ever compute the proper low-spin wavefunction, and so only the energy is generally able to be extracted. Second, the results become highly functional dependent. While all systems demonstrate some density functional dependence, spin-coupled complexes are intrinsically more sensitive because the percentage of exact exchange directly affects the relative energies of the high spin and broken symmetry states \cite{reiherReparameterizationHybridFunctionals2001, reiherTheoreticalStudyFe2002}. 
Finally, DFT doesn't offer a path toward systematic improvements, making it difficult to compare results. 

Because of these reasons, multireference methods like CASSCF and CASPT2 are often used to model exchange coupled systems. However, the associated computational cost limits the active space size, making it difficult to converge results to the point where quantitative comparison to experiment is possible. Often one finds that dynamical correlation (involving interactions with non-magnetic orbitals) has a significant impact on the value of $J$, generally strengthening the antiferromagnetic interactions. As such, DMRG has emerged as the standard benchmark method for computing exchange coupling constants in organometallic compounds 
\cite{pantazisMeetingChallengeMagnetic2019,
szalayTensorProductMethods2015,
baiardiDensityMatrixRenormalization2020,
schollwockDensitymatrixRenormalizationGroup2011,
woutersDensityMatrixRenormalization2014,
eriksenGroundStateElectronic2020,
olivares-amayaAbinitioDensityMatrix2015a,
sharmaLowenergySpectrumIronsulfur2014, 
sharmaQuasidegeneratePerturbationTheory2016a,
harrisInitioDensityMatrix2014},
although if only the value of $J$ is needed, efficient approaches that combine spin-flip methods have also been useful \cite{mayhallComputationalQuantumChemistry2014, mayhallComputationalQuantumChemistry2015a, mayhallSpinflipNonorthogonalConfiguration2014a,pokhilkoEffectiveHamiltoniansDerived2020}.

\begin{figure}
    \centering
    \includegraphics[width=\linewidth]{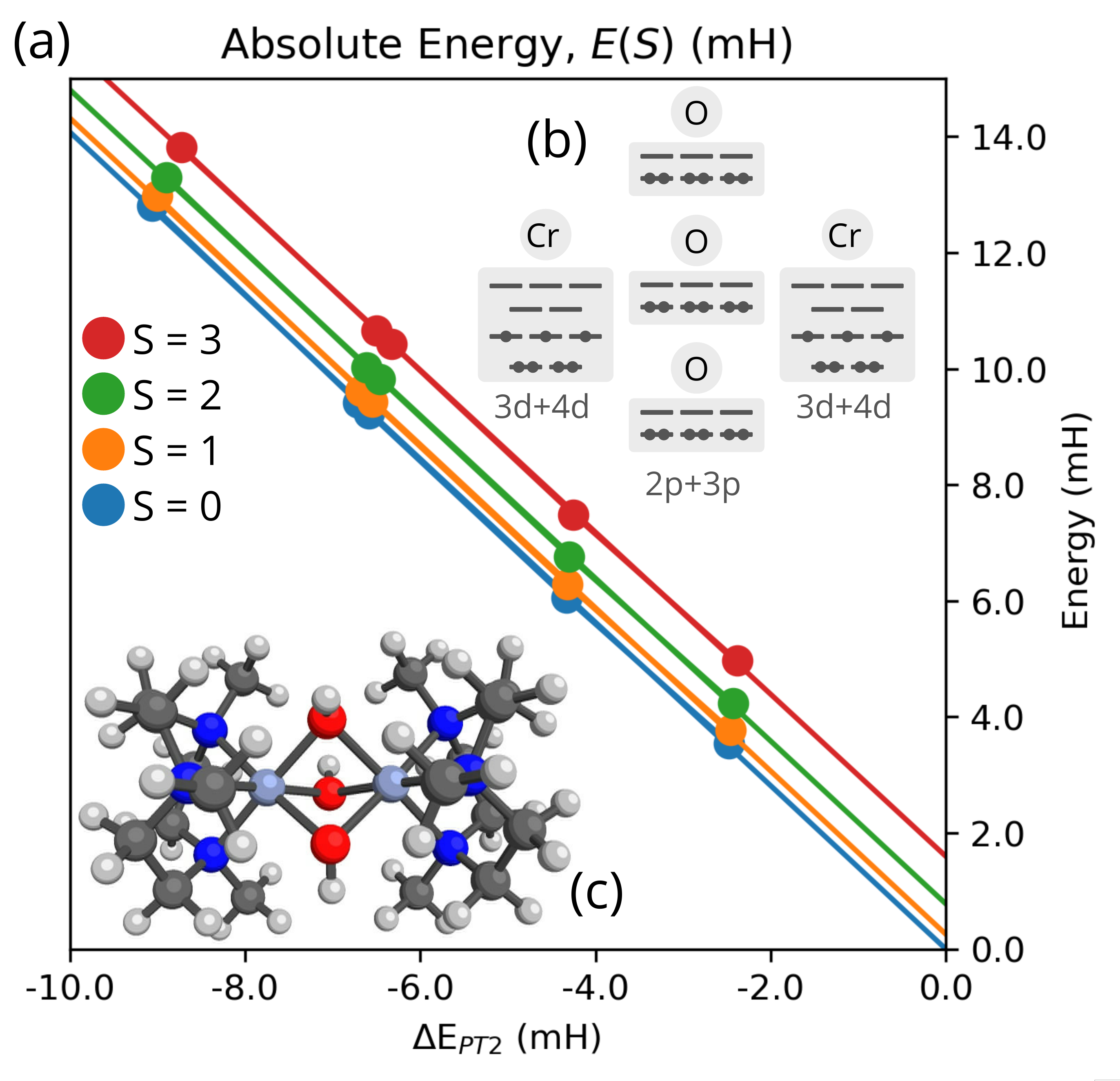}
    \caption{Convergence and extrapolation of Cr$_2$ low-energy spectra for the
    def2-SVP basis and a (32e, 38o) active space.
    (a) Plot of the TPSCI variational energy as a function of the computed PT2 correction. Solid line is the linear fit.
    Units in milliHartree. $M$=100.
    (b) Clustering of the 38 orbital active space. 
    (c) Molecular structure}
    \label{fig:cr2_extrap}
    
\end{figure}

In this subsection, we present the first application of TPSCI to a transition metal compound,
a tris-hydroxy-bridged Cr(III) dimer,  [L$_2$Cr(III)$_2$($\mu$-OH)$_3$]$^{+3}$ (L = N,N',N''-trimethyl-1,4,7-triazacyclononane (Fig. \ref{fig:cr2_extrap}(c)). 
which has a $J$ value that was experimentally fit to a value of $-66$ cm$^{-1}$ \cite{niemannNewStructureMagnetism1992}. 
Recently, Pantazis studied this system using  DMRG to solve the low-energy states in a large orbital active space of up to (30e, 22o) \cite{pantazisMeetingChallengeMagnetic2019},
calculating an exchange coupling constant of $-23.9$ cm$^{-1}$. 

\subsubsection{Active Space Selection} 
For our calculations, we extend the size of the active space by including the orbitals which overlap most strongly with the 3d and 4d orbitals on each Cr center, as well as the 2p and 3p oxygen orbitals on each bridging OH$^{-1}$ ligand, leading to an overall orbital active space of 38 orbitals. Our active space was obtained by first optimizing the ROHF wavefunction for the high-spin heptet state. We then define a set of atomic orbitals for which we would like to span as closely as possible without destroying the ROHF reference. For this system, we take as our projection space ($\mu_A$), the Cr 3d and 4d atomic orbitals, and the bridging O 2p and 3p orbitals, leading to a total of 38 orbitals. We then separately project the doubly occupied ($i$), singly occupied ($s$), and virtual ($a$) orbital spaces onto this AO subspace, providing matrices $C_{\mu_A, i}, C_{\mu_A, s}, C_{\mu_A, a}, $. We then perform separate SVD's on each projected subspace, and keep the largest singular vectors from each orbital space. As such, we start with 38 atomic orbitals and end up with 38 molecular orbitals. Following this automated procedure produced an active space consisting of 13 doubly occupied, 6 singly occupied (the full ROHF open shell space), and 19 virtual orbitals, leading to a (32e, 38o) active space. While this is not the only way to yield an active space, it was convenient for our purposes as the resulting orbital active spaces are already localized to our target system. In the future, more extensive tests will be performed for automating the construction of localized active spaces. The active orbitals are shown in the supplementary information.

\subsubsection{Clustering}
The 38 active orbitals described above were then organized into 5 clusters, as depicted in Fig. \ref{fig:cr2_extrap}(b).  Here, each Cr atom cluster defined a local (7e, 10o) active space, and each oxygen a (6e, 6o) local active space. 
For each cluster, a local many-body basis was constructed from the $M$ lowest energy states for each sector of Fock space which contained up to $N_I \pm \delta_{elec}$ number of electrons, where we set $\delta_{elec}=3$ for these calculations. For example, each Cr cluster has a basis of up to $M$ states obtained by diagonalizing the CMF Hamiltonian for each of the following active spaces: (4e,10o), (5e,10o), (6e,10o), (7e,10o), (8e,10o), (9e,10o), and (10e,10o). 
Similarly, each bridging $\text{OH}^{-1}$ ligand had 7 different active spaces centered at (6e, 6o).\footnote{For each active space, we solve for the lowest $M$ states in the lowest $M_s$-subspace, and then generate the higher $M_s$ vectors by directly applying $\hat{S}^{\pm}$. This is done to, not only reduce the cost, but also to ensure that every TPS in our basis can be properly spin-adapted to form proper eigenstates of $\hat{S}^2$, as first noted in Ref. \cite{braunscheidelGeneralizationTensorProduct2023a}.}

As is commonly done with selected CI approaches, significantly improved approximations to the energy can obtained by performing a series of selected CI calculations with varying thresholds and extrapolating to the zero error limit, which is taken as an approximation to full CI. While more sophisticated extrapolation schemes have been proposed \cite{burtonRationaleExtrapolationProcedure2024},
we use the common approach of assuming a linear relationship between the variational energies and the PT2 correction \cite{holmesExcitedStatesUsing2017}. 
In Fig. \ref{fig:cr2_extrap}(a), we plot the variational TPSCI energy of the 4 lowest eigenstates as a function of the PT2 correction to each state. By extrapolating this linear relationship to zero PT2 correction, we can obtain an estimate of the exact eigenvalues of the Hilbert space defined by $M$. In Fig. \ref{fig:cr2_extrap}(a), we show the extrapolation for the $M=100$ calculations.

\begin{table}
    \centering
    \begin{tabular}{c|ccc}
    \hline\hline
         & $J$ (S0,S1)& $J$ (S1,S2)& $J$ (S2,S3)\\\hline
        & \multicolumn{3}{c}{TPSCI}  \\
        $M=100$ & -25.4 & -25.7 & -26.6\\
        $M=200$ & -26.6 & -27.0 & -27.7\\
        & \multicolumn{3}{c}{TPSCI+PT2}  \\
        $M=100$ & -26.7 & -27.3 & -28.3\\
        $M=200$ & -28.3 & -28.9 & -29.9\\
        & \multicolumn{3}{c}{Extrapolated}  \\
        $M=100$ & -28.0 & -28.7 & -30.0\\
        $M=200$ & -29.3 & -30.3 & -31.3\\  
    \hline\hline
\end{tabular}
    \caption{Exchange coupling constants (cm$^{-1}$) for Cr$_2$ compound with def2-SVP basis, and (32e, 38o) active space.
    $J$ refers to $H=-2J\hat{S}_1\cdot\hat{S}_2$.
    $J$(S0,S1) denotes which spin states are used to computed $J$ via Land\'e rule.
    ``TPSCI'' refers to the best variational energy obtained, using $\epsilon_{cipsi}=\num{2e-4}$.
    ``TPSCI+PT2'' is the best variational energy plus the state-specific PT2 correction.
    ``Extrapolated'' uses differences between the extrapolated energies.
    $M$ is the maximum number of cluster states computed for each cluster Fock sector. 
    The dimension of the sparse variational TPSCI subspace is 97357 for M=100 and 127493 for M=200. 
    }
    \label{tbl:cr2_j}
\end{table}

In order to compute $J$,  we can use the Land\'e interval rule derived from the energy spectrum of a two-site Heisenberg model, $J = (E(S-1)-E(S))/2S$. After computing the lowest energy S=0, 1, 2, and 3 spin states, we could use any of the gaps to compute $J$. If the ab initio system were to be perfectly described by the Heisenberg model, then the computed $J$ value would be independent of the particular energy gaps we were to choose. 
However, the Heisenberg model is rarely, exact, and so we can partially quantify how approximate the model is by comparing $J$ values computed with different energy gaps. We list the various $J$ values in Table \ref{tbl:cr2_j}, using either the best variational TPSCI energies, the TPSCI+PT2 corrected energies, or the extrapolated energies. Results for both $M=100$ and $M=200$ are included.\footnote{We have also explored directly extrapolating the $J$ values themselves, and found the results to be essentially the same.}

Inspecting first the effect of energy extrapolation, we find that the extrapolated $J$ values are larger than the TPSCI+PT2 values by only around 1 $cm^{-1}$, and that doubling the size of $M$ from 100 to 200 only increases the $J$ value by another 1 $cm^{-1}$, despite the fact that this also increases the dimension of the total accessible Hilbert space significantly from $\num{2.7e14}$ to $\num{6.1e15}$.
Compared to the recent (30e, 22o)  DMRG-SCF calculations which yielded a $J$ value of $-23.9$ cm$^{-1}$ \cite{pantazisMeetingChallengeMagnetic2019}, our computed values are slightly larger, in good agreement with the reported CASSCF(6e,10o)-NEVPT2 values of $-31.8$ $cm^{-1}$ \cite{pantazisMeetingChallengeMagnetic2019}.

\begin{figure*}
    \centering
    \includegraphics[width=\linewidth]{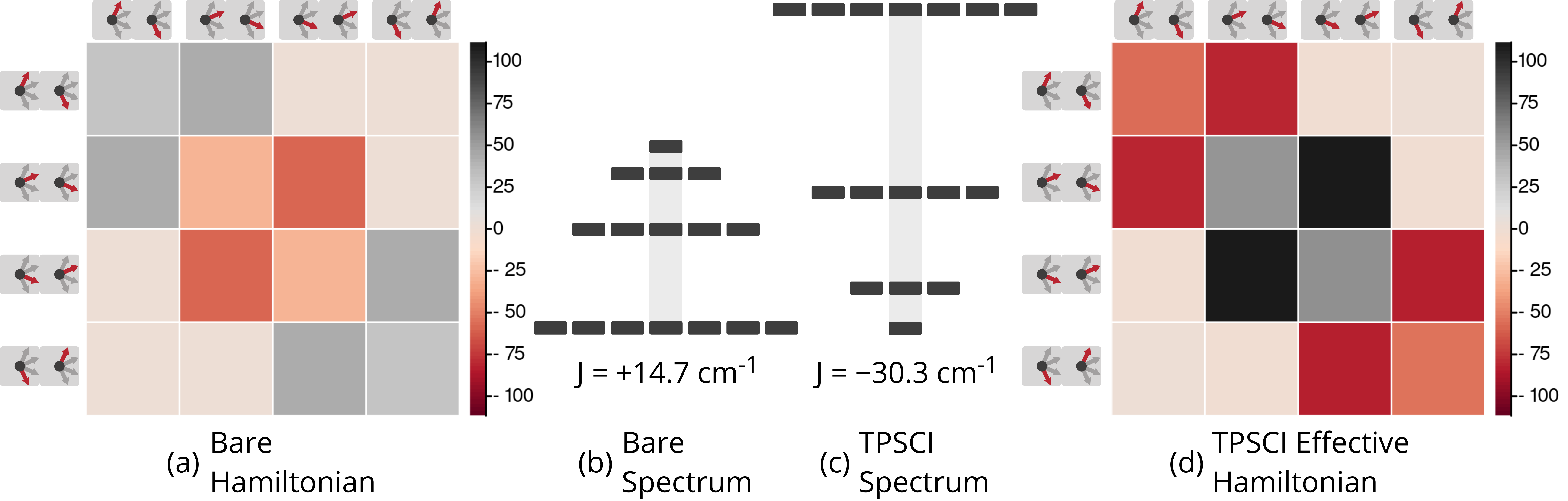}
    \caption{Bare and Bloch Effective Hamiltonians for Cr$_2$(III) complex. 
    (32e, 38o) active space. Units of $J$ in cm$^{-1}$.
    (a) Hamiltonian matrix in the basis of $M_s=0$ tensor products of local spin states where all the bridging ligands are in singlet states, and the Cr centers are in $\ket{S=\tfrac{3}{2}, M_s=\pm\tfrac{3}{2}}$
    (b) Energy spectrum of bare Hamiltonian in local $S=\tfrac{3}{2}$ basis.
    (c) Energy spectrum of extrapolated TPSCI results.
    (e) Bloch Effective Hamiltonian obtained by projecting TPSCI wavefunctions onto local $S=\tfrac{3}{2}$ basis.}
    \label{fig:cr2_bloch}
\end{figure*}
\subsubsection{Bloch Effective Hamiltonian}
In order to further analyze the results listed in Table \ref{tbl:cr2_j}, we compute a Bloch effective Hamiltonian, providing access to the individual effective (dressed) interactions between the various spin microstates that lead to the low-energy spectrum. A qualitative description of this complex assigns each Cr center a well-defined oxidation state (III) and spin state (S=$\tfrac{3}{2}$). As such, the low-energy spectrum is expected to be dominated by the 16 spin configurations that form a basis for the S = 3, 2, 1, and 0 states, providing a clear definition for our model space. However, because the Hamiltonian preserves spin, we can restrict our focus to only the global $M_s=0$ subspace, and take our model space to be the corresponding 4 dimensional subspace. In Fig. \ref{fig:cr2_bloch}, we plot both the bare Hamiltonian (Fig. \ref{fig:cr2_bloch}(a)) and the Bloch-effective Hamiltonian  (Fig. \ref{fig:cr2_bloch}(d))  in the model space. The corresponding low-energy spectra are provided in Figs. \ref{fig:cr2_bloch}(b) and \ref{fig:cr2_bloch}(c), respectively. 

There are two main features of the effective Hamiltonian that emerge from the implicit inclusion of the external space correlation: the spin-coupling interactions have their signs flipped, and their magnitudes increased. The result of this is that the system changes from ferromagnetic to antiferromagnetic coupling and the gaps between the spin states are approximately doubled. 

This qualitative result is consistent with a recent study (Ref. \cite{pandharkarLocalizedActiveSpaceState2022}) where $J$ values for a similar dichromium (III) complex were computed using the vLASSCF-SI method. 
Because the vLASSCF method works in a similar TPS basis, they were able to evaluate the impact of explicitly including some charge transfer configurations.
They too found that this affected a qualitative change in the sign of $J$, switching from ferromagnetic to anti-ferromagnetic.

\begin{table}
    \centering
    \begin{tabular}{l|cccc}
    \hline\hline
Root &	1 &	2 &	3 &	4  \\
$\ev{\hat{S}^2}$ &	0.000 &	2.000 &	6.000 &	12.000  \\
\hline
$\ev{\hat{N}_{Cr}}$ &	7.010 &	7.010 &	7.010 &	7.010  \\
var($\hat{N}_{Cr}$) &	0.019 &	0.018 &	0.016 &	0.014  \\
cov($\hat{N}_{Cr},\hat{N}_{Cr}$) &	-0.004 &	-0.004 &	-0.002 &	0.000  \\
\hline
$\ev{\hat{S}^z_{Cr}}$ &	0.019 &	0.013 &	0.003 &	0.003  \\
var($\hat{S}^z_{Cr}$) &	1.248 &	2.042 &	1.246 &	0.453  \\
cov($\hat{S}^z_{Cr},\hat{S}^z_{Cr}$) &	-1.245 &	-2.039 &	-1.242 &	-0.449  \\
\hline
$\ev{\hat{S}^2_{Cr}}$ &	3.741 &	3.742 &	3.745 &	3.750  \\
var($\hat{S}^2_{Cr}$) &	0.064 &	0.061 &	0.056 &	0.047  \\
cov($\hat{S}^2_{Cr},\hat{S}^2_{Cr}$) &	0.017 &	0.014 &	0.008 &	-0.001 \\
\hline\hline
\end{tabular}
    \caption{Local expectation values for Cr$_2$(III) complex.
    Variance is computed as: $\mathrm{var}\left(\hat{O}_{Cr}\right)=\ev{\hat{O}_{Cr}^2}-\ev{\hat{O}_{Cr}}^2$. 
    Covariance is computed as: $\mathrm{cov}\left(\hat{O}_{Cr}\right)=\ev{\hat{O}_{Cr_A}\hat{O}_{Cr_B}}-\ev{\hat{O}_{Cr_A}}\ev{\hat{O}_{Cr_B}}$. 
    }
    \label{tbl:cr2_corr}
\end{table}

In order to further analyze this Cr$_2$(III) system, we can compute cumulants of local observables, as mentioned above \cite{casanovaQuantifyingLocalExciton2016a, luzanovQuantifyingChargeResonance2015a}.
In Table \ref{tbl:cr2_corr}, we list the expectation values, the variances, and the covariances of a few operators local to the Cr$_2$(III) centers, including local particle number, $\hat{S}^z_{Cr}$, and $\hat{S}^2_{Cr}$.  We also include the global $\hat{S}^2$ because our basis is not spin-adapted, so there is potential for spin-contamination, although our calculations are converged tightly enough to reduce spin-contamination to the 4th decimal place.

Considering first the local particle number values, we find that the average number of electrons is quite consistent across the different spin states, staying just barely above 7 electrons (which corresponds to a Cr(III) oxidation state, with 2 doubly occupied ligand orbitals).  However, the fluctuations in the oxidation state noticeably depend on spin state, increasing as the global spin is decreased.  This is easily understood as a consequence of the super-exchange mechanism, whereby coupling to electron transfer states stabilizes the low-spin states relative to the high spin states. 
Again consistent with that seen in Ref. \cite{pandharkarLocalizedActiveSpaceState2022}, and the more general treatment of formal magnetic interactions from Malrieu and coworkers \cite{malrieuMagneticInteractionsMolecules2014}. 
For the global singlet state, the statistical correlation between the oxidation state fluctuations on the two Cr centers is only 21.1\%, indicating that the majority of the oxidation state fluctuations are due to electron exchanges with the bridging ligands. 

Because both Cr centers are re-coupled into global eigenvectors of  $\hat{S}^2$, local $\hat{S}^z_{Cr}$ is no longer a good quantum number, and becomes maximally uncertain. 
In fact, we can further test how closely the system follows Heisenberg-Dirac-von Vleck physics by comparing the analytic values of the local $S^z_{Cr}$ variance using the Clebsch-Gordon coefficients for a product of two $S=\tfrac{3}{2}$ spins.  For the singlet, triplet, quintet, and heptet states, the analytic $\textrm{var}\left(\hat{S}^z_{Cr}\right)$ values are $-1.25$,  $-2.05$, $-1.25$, and $-0.45$, respectively. Our computed correlations are only slight different from these analytical values: $-1.245, -2.039, -1.242, -0.449$, further indicating good consistency with the Heisenberg model. 

Inspecting the local $\hat{S}^2_{Cr}$ values, we see a complementary picture to that provided by the particle number fluctuations. 
As the global spin is decreased, the local $S^2$ values also tend to decrease, while the variance increases. This is consistent with the superexchange mechanism stabilizing global low-spin states by coupling to charge transfer states. 
When an electron transfers from one Cr to the another, the number of unpaired electrons decreases. As a result, the local $S^2_{Cr}$ values decrease, and develop a positive covariance between the centers. 

\begin{figure*}
    \centering
    \includegraphics[width=1\linewidth]{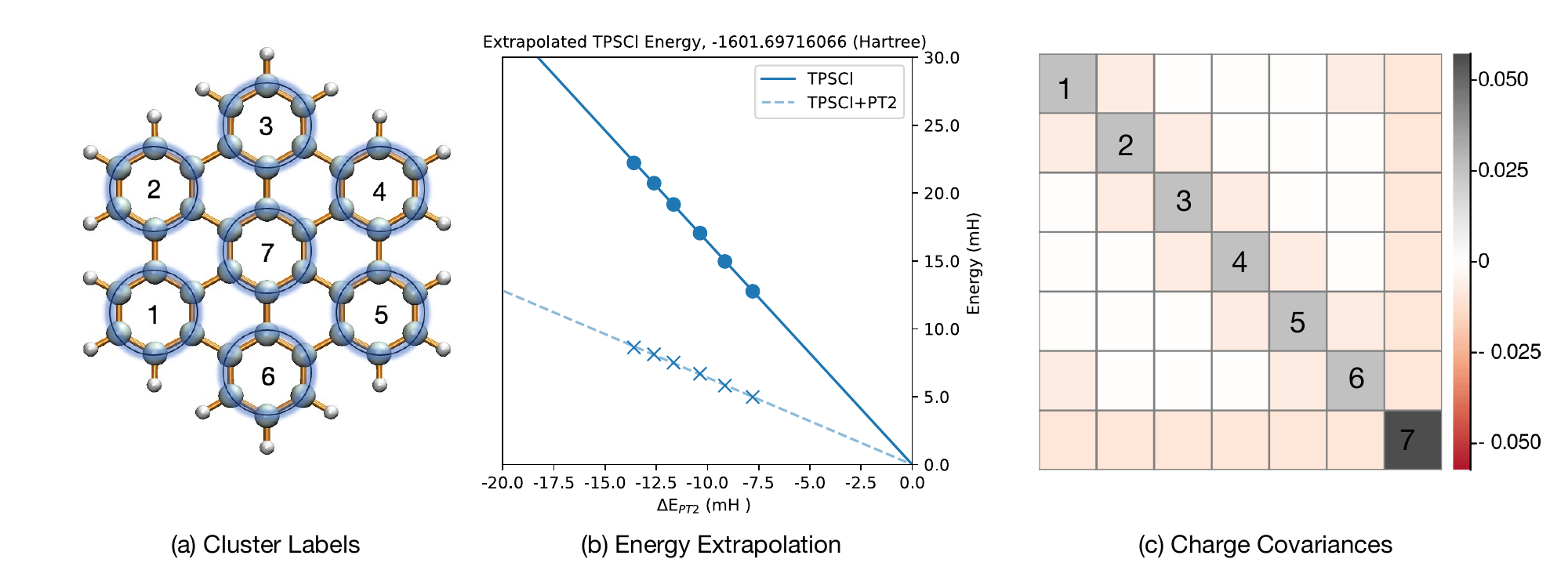}
    \caption{Hexabenzocoronene. (a) Molecular structure and cluster indices. Active space (42e, 42o) includes all $\pi$ orbitals.
    (b) Extrapolation of energy of the singlet ground state. Units of milli-Hartree.
    (c) Charge covariance matrix. cov$\left(\hat{N}_I, \hat{N}_J\right)$}
    \label{fig:hbc_extrap}
\end{figure*}

\subsection{Conjugation in 2D}\label{sec:hbc}
In the earlier sections, the 
tetracene tetramer (Sec. \ref{sec:sf}) served as an example of a completely non-bonded systems, which is clearly quite easily clusterable. 
In Sec. \ref{sec:cr2}, we demonstrated that the concepts of oxidation state and local spin allowed us to treat the Cr$_2$(III) complex in a clustered representation.
In contrast, conjugated $\pi$-systems are characterized primarily by the highly delocalized nature of the electronic structure. In this section, we investigate the ability to compute and analyze the full $\pi$ active space for a large delocalized $\pi$ system.

Hexabenzocoronene (C$_{42}$H$_{18}$) is a polycyclic aromatic hydrocarbon (shown in Fig. \ref{fig:hbc_extrap}(a)) where an additional benzene ring is fused to the outside of a central coronene ring. The $\pi$ electrons are delocalized across the entire molecule which contributes to its unique electronic properties. As we have seen previously \cite{abrahamSelectedConfigurationInteraction2020c}, the delocalized hexabenzocoronene system serves as a nice edge case for evaluating the ability of TPSCI to provide both accurate and insightful results for systems that are not obviously clusterable.

\subsubsection{Active Space selection and clustering}
For these results, we have considered the full $\pi$-system orbital active space (42 orbitals consisting of the $2p_z$ orbitals on each carbon) using cc-pVDZ basis set\cite{dunning_gaussian_1989}.
Viewing hexabenzocoronene as a collection of seven benzene rings, we partition the 42 orbitals into 7 clusters of 6 orbitals. A depiction of this clustering is shown in Fig. \ref{fig:hbc_extrap}(a). 
The geometry is optimized at the B3LYP/cc-pVDZ level of theory. The cMF optimization (including orbital rotations between clusters) is performed using our open-source Julia package \href{https://github.com/nmayhall-vt/ClusterMeanField.jl}{ClusterMeanField.jl}\cite{mayhall_nicholas_clustermeanfield_nodate}.
For each cluster, a local many-body basis was constructed using the embedded Schmidt truncation (EST) approach, where we define the cluster basis as the singular vectors of FCI ground state on the cluster plus an orbital bath.  We discarded Schmidt vectors with singular values smaller than a threshold value of $1e^{-4}$. Detailed analysis of EST approach has been done in our recent TPSCI paper\cite{abrahamSelectedConfigurationInteraction2020c}.

\subsubsection{Convergence of TPSCI ground state}
In Fig. \ref{fig:hbc_extrap}(b), we plot the extrapolation of the TPSCI variational energy as a function of the PT2 correction. We use $\epsilon_{\text{FOIS}}=1 \times 10^{-6}$ (threshold on the external space couplings), and tightest $\epsilon_{\text{CIPSI}}=1.5 \times 10^{-4}$ through the bootstrapping HOSVD approach for this calculation\cite{braunscheidelGeneralizationTensorProduct2023a}. Here we see that our ground state TPSCI+PT2 energy is only about 5 mH away from the extrapolated result. Further, this was with a variational dimension of only around 114k. The extrapolated total energy of the molecule is $-1601.6971 \pm 0.0001$ au. It's important to emphasize that the uncertainty here is due to the linear fit and does not imply a variational guarantee. If we \textit{knew} that the relationship was indeed linear over the full range, then our energy would be correct to 0.1 mH.

We note that the performance of TPSCI on such delocalized $\pi$ systems depends significantly on the topology of the molecule. For the complex considered here, there is a well defined Clar's structure that suggests a unique clustering. We expect this to be key to achieving accurate solutions. In contrast, our recent work \cite{abrahamSelectedConfigurationInteraction2020c} revealed that $\pi$ systems without a well-defined clustering into a Clar's structure (such as coronene) is significantly slower to converge. We plan to explore this topic in the future for a more extensive set of polyaromatic hydrocarbons.


\subsubsection{Correlations in between clusters}\label{sec:correlations}
In Fig. \eqref{fig:hbc_extrap}(c), the charge covariances between the clusters are depicted as a heatmap, with each row/column corresponding to a given cluster labeled by the number on the diagonal. 
Looking first at the diagonal of the matrix (the charge variances), we see that the outside clusters (1-6) all have equivalent charge fluctuations, while the central benzene unit has significantly larger fluctuations in the ground state. Because the variance quantifies the uncertainty in the number of electrons in a given cluster, each of the outer and inner clusters have  $6.0\pm0.5$ and $6.0 \pm 0.7$, numbers of electrons, respectively (using a 3$\sigma$ confidence interval). 
Considering next the off-diagonal matrix elements, we see that all nearest neighbor cluster couplings are approximately the same. 
Assuming two-body correlations dominate, this means that the central carbon should have a variance that is about twice that of the outer clusters, just based on the fact that it has twice as many nearest neighbors, which is consistent with the observed results. 
Because \ small differences are difficult to see in the heatmap, we have listed the unique covariance quantities in Table \ref{tbl:hbc_corr}.
Very similar results exist for the $\hat{S}^z_I$ correlations, as can be seen in Fig. S2 in the supplementary information.

By considering the wavefunction directly, we notice that about 89.6\% of the wavefunction is attributable to TPS's that have $6\alpha$ and $6\beta$  in each cluster, whereas 9.1\% is due to charge transfer configurations, and 1.3\% is due to local spin-flip configurations. 

We also note, that in Table \ref{tbl:hbc_corr}, we see stronger particle number and spin correlations between benzenes connected in the para position than between those with meta connections, despite being further in distance, indicating a slight directing effect of the central benzene.  
Although the opposite is seen with the cluster excitation, $\hat{Q}_I$, correlations. 


\begin{table}[h!]
\centering
\caption{Unique cluster correlations in hexabenzocoronene.
$\hat{N}_I$ is the number operator for cluster $I$. 
$\hat{S}^z_I$ is the spin magnetization operator for cluster $I$.
$\hat{Q}_I$ is the projector onto the orthogonal complement of the cMF ground state for cluster $I$.
Cluster pair indices correspond to the labeling in Fig. \ref{fig:hbc_extrap}(a) with the description of interaction type in parenthesises.}
\label{tbl:hbc_corr}
\begin{tabular}{l | ccc}   
\hline\hline
Cluster Pair & cov$\left(\hat{N}_i,\hat{N}_j\right)$& cov$\left(\hat{S}^z_i,\hat{S}^z_j\right)$& cov$\left(\hat{Q}_i,\hat{Q}_j\right)$\\\hline
1,1 (outer) & 0.02551& 0.00925& 0.03307
\\
7,7 (inner) & 0.05729& 0.02103& 0.06867
\\
1,2 (nearest neighbor)& -0.00759& -0.00276& 0.01079
\\
1,3 (meta) & -0.00018&  -0.00005& -0.00010
\\
1,4 (para) & -0.00040& -0.000134& -0.00006
\\
1,7 (outer-inner)& -0.00956& -0.00350& 0.01342\\\hline
\end{tabular}
\end{table}

Although the particle number covariance between each pair of neighboring clusters is negative (indicating charge transfer), the $\hat{S}^2_I$ correlations are positive between neighboring clusters.  This is consistent with the description of charge correlation. When an electron from a cluster hops into another cluster, then a doublet state will be formed in both of the clusters. One cluster will be one electron deficient giving rise to a cationic doublet state and the extra electron forms an anionic doublet state in the neighboring cluster. Consequently, when one cluster is in a doublet state, its neighbors have a higher probability of also being found in a doublet state, making the correlation positive. 
The entanglement between clusters leads each to acquire a non-zero average $\hat{S}^2_I$ value. 
The outer clusters have total spin of $0.034\pm 0.55$, whereas the central cluster as a  local $S^2_I$ value of $0.078 \pm 0.84$, where uncertainty is given as 3$\sigma$.

\section{Conclusions}
In this paper, we have explored the ability of TPSCI to provide accurate yet interpretable approximations to FCI on relatively large orbital active spaces. 
Because TPSCI works in a basis consisting of products of local FCI states, the more separable a system is, the easier it should be to simulate. 
As such, in this paper, we consider three example systems, which range in the degree of separability: (i) a completely non-bonded tetramer of tetracene molecules, (ii) a more strongly interacting dichromium organometallic complex which, while bonded, is still characterized with local quantities like oxidation state, and (iii) a completely delocalized $\pi$ system of hexabenzocoronene.  

For the dichromium example, this was the first TPSCI calculation applied to open-shell biradical systems. We found that TPSCI was able to compute exchange coupling constants that are larger in magnitude (presumably more accurate), than recent DMRG calculations. 

For each of these systems we characterized the resulting wavefunctions using quantities that are easily accessible from the TPS basis.
By leveraging the natural diabatic character of the TPS basis, we are able to easily construct Bloch effective Hamiltonians, which provide quantitative relationships between physically relevant degrees of freedom. 
This provided access to quantities such as the effective coupling between the bright states and the multiexcitonic states, $\bra{\mathrm{S}_1}\hat{H}^{eff}\ket{^1\mathrm{TT}}$, which implicitly includes the downfolded effects from charge transfer couplings, which are substantially enhanced compared to the direct coupling. 

We additionally used correlation functions of quantities like particle number, spin, and cluster excitation to provide a more detailed analysis of the various variational TPSCI eigenstates, and ultimately compare the results to inspection of the wavefunction itself which is particularly interpretable due to the diabatic nature of the basis. 
This work, helps lay out approaches for extracting more insight from TPSCI wavefunctions (and other TPS based wavefunctions) in the future.

\section{Acknowledgments}
This work was supported by the National Science Foundation (Award No. 1752612). NMB also acknowledges the support provided by the Graduate School Doctoral Assistantship from the Department of Chemistry at Virginia Tech. 

\providecommand{\latin}[1]{#1}
\makeatletter
\providecommand{\doi}
  {\begingroup\let\do\@makeother\dospecials
  \catcode`\{=1 \catcode`\}=2 \doi@aux}
\providecommand{\doi@aux}[1]{\endgroup\texttt{#1}}
\makeatother
\providecommand*\mcitethebibliography{\thebibliography}
\csname @ifundefined\endcsname{endmcitethebibliography}
  {\let\endmcitethebibliography\endthebibliography}{}

\end{document}


\beginsupplement
\title{Supporting Information: Accurate and Interpretable Representation of  Correlated Electronic Structure via Tensor Product Selected CI}

\author{Nicole M. Braunscheidel}
\affiliation{Department of Chemistry, Virginia Tech, Blacksburg, VA 24060, USA}
\author{Arnab Bachhar}
\affiliation{Department of Chemistry, Virginia Tech, Blacksburg, VA 24060, USA}

\author{Nicholas J. Mayhall}
\email{nmayhall@vt.edu}
\affiliation{Department of Chemistry, Virginia Tech, Blacksburg, VA 24060, USA}

\maketitle

\section{Singlet Fission Tetracene Tetramer}
\begin{table}[h!]
\label{SI:tt}
\begin{center}
\caption{Results for all 31 eigenstates of tetracene tetramer with associated labels based on expectation values of the $\hat{S}^2$ operator $\expval{\hat{S}^2}$, variational excitation energies ($\omega_\text{Var}$), PT2 energy corrections ($\omega_\text{PT2}$) for excitation energies in eV, and percentage of charge transfer ($\%$ CT) for all 31 eigenstates in the TPSCI wavefunction. With $\epsilon_{cipsi}=4e-4$ resulting in a variatonal dimension of 128,614.}
\begin{tabular}{|c|c|c|c|c|c|}
\hline 
State & Label & $\expval{\hat{S}^2}$ & $\omega_\text{Var}$ & $\omega_\text{PT2}$ &  \% CT\\ \hline
1	&	S$_0$	&	0.000	&	0.000	&	0.000	&	0.15	\\
2	&	T$_1$	&	2.000	&	1.813	&	0.004	&	1.49	\\
3	&	T$_1$	&	2.000	&	1.834	&	0.003	&	0.67	\\
4	&	T$_1$	&	2.000	&	1.848	&	0.003	&	0.68	\\
5	&	T$_1$	&	2.000	&	1.861	&	0.003	&	0.44	\\
6	&	${}^1(\mathrm{TT})$	&	0.000	&	3.635	&	0.011	&	4.55	\\
7	&	${}^1(\mathrm{TT})$	&	0.002	&	3.645	&	0.010	&	3.09	\\
8	&	${}^3(\mathrm{TT})$	&	2.000	&	3.647	&	0.009	&	2.26	\\
9	&	${}^5(\mathrm{TT})$	&	5.998	&	3.651	&	0.009	&	1.82	\\
10	&	${}^1(\mathrm{TT})$	&	0.000	&	3.653	&	0.010	&	4.49	\\
11	&	${}^3(\mathrm{TT})$	&	2.000	&	3.654	&	0.010	&	3.41	\\
12	&	${}^1(\mathrm{TT})$	&	0.001	&	3.664	&	0.009	&	2.89	\\
13	&	${}^3(\mathrm{TT})$	&	2.000	&	3.665	&	0.009	&	3.29	\\
14	&	${}^3(\mathrm{TT})$	&	2.000	&	3.675	&	0.009	&	2.33	\\
15	&	${}^5(\mathrm{TT})$	&	6.000	&	3.682	&	0.009	&	1.52	\\
16	&	${}^5(\mathrm{TT})$	&	5.996	&	3.688	&	0.008	&	1.10	\\
17	&	${}^5(\mathrm{TT})$	&	5.591	&	3.694	&	0.007	&	0.84	\\
18	&	${}^3(\mathrm{TT})$	&	1.366	&	3.695	&	0.007	&	1.05	\\
19	&	${}^1(\mathrm{TT})$	&	1.100	&	3.695	&	0.007	&	1.06	\\
20	&	${}^5(\mathrm{TT})$	&	5.944	&	3.698	&	0.007	&	0.76	\\
21	&	${}^1(\mathrm{TT})$	&	0.002	&	3.707	&	0.008	&	1.35	\\
22	&	${}^3(\mathrm{TT})$	&	1.999	&	3.709	&	0.008	&	1.25	\\
23	&	${}^5(\mathrm{TT})$	&	5.999	&	3.715	&	0.008	&	0.68	\\
24	&	S$_1$	&	0.000	&	3.809	&	0.010	&	12.25	\\
25	&	S$_1$	&	0.000	&	3.884	&	0.007	&	3.49	\\
26	&	T$_2$	&	2.000	&	3.900	&	0.005	&	2.42	\\
27	&	T$_2$	&	2.000	&	3.916	&	0.004	&	1.30	\\
28	&	S$_1$	&	0.000	&	3.920	&	0.008	&	7.50	\\
29	&	T$_2$	&	2.000	&	3.924	&	0.005	&	2.33	\\
30	&	T$_2$	&	2.000	&	3.928	&	0.005	&	2.58	\\
31	&	S$_1$	&	0.000	&	3.952	&	0.007	&	5.91	\\ \hline
\end{tabular}
\end{center}
\end{table}

\begin{table}[h!]
\label{SI:heff_bare}
\begin{center}
\caption{Bare Hamiltonian to show singlet coupling strengths between $\ket{\mathrm{S}_1}$ and $\ket{^1\mathrm{TT}}$ in meV.}
\begin{tabular}{|c|c|c|c|c|c|c|c|c|c|c|c|}
\hline 
State	&	S$_0$	&	S$_1$	&	S$_1$	&	S$_1$	&	S$_1$	&	1,2	&	1,4	&	2,3	&	1,3	&	2,4	&	3,4	\\ \hline
S$_0$	&	0.00	&	0.00	&	0.00	&	0.00	&	0.00	&	-0.84	&	-0.72	&	-1.38	&	-0.88	&	-0.09	&	-0.01	\\
S$_1$	&	0.00	&	0.00	&	-11.75	&	-38.45	&	9.45	&	-0.22	&	-0.43	&	0.00	&	-0.05	&	0.00	&	0.00	\\
S$_1$	&	0.00	&	-11.75	&	0.00	&	-10.15	&	40.67	&	-0.68	&	0.00	&	-0.94	&	0.00	&	-0.03	&	0.00	\\
S$_1$	&	0.00	&	-38.45	&	-10.15	&	0.00	&	-8.07	&	0.00	&	0.00	&	-0.59	&	0.01	&	0.00	&	0.00	\\
S$_1$	&	0.00	&	9.45	&	40.67	&	-8.07	&	0.00	&	0.00	&	1.06	&	0.00	&	0.00	&	-0.01	&	0.00	\\
1,2	&	-0.84	&	-0.22	&	-0.68	&	0.00	&	0.00	&	0.00	&	-0.05	&	0.32	&	1.44	&	1.12	&	0.00	\\
1,4	&	-0.72	&	-0.43	&	0.00	&	0.00	&	1.06	&	-0.05	&	0.00	&	0.00	&	0.01	&	-0.08	&	0.32	\\
2,3	&	-1.38	&	0.00	&	-0.94	&	-0.59	&	0.00	&	0.32	&	0.00	&	0.00	&	-0.08	&	0.01	&	-0.05	\\
1,3	&	-0.88	&	-0.05	&	0.00	&	0.01	&	0.00	&	1.44	&	0.01	&	-0.08	&	0.00	&	0.00	&	1.12	\\
2,4	&	-0.09	&	0.00	&	-0.03	&	0.00	&	-0.01	&	1.12	&	-0.08	&	0.01	&	0.00	&	0.00	&	1.44	\\
3,4	&	-0.01	&	0.00	&	0.00	&	0.00	&	0.00	&	0.00	&	0.32	&	-0.05	&	1.12	&	1.44	&	0.00	\\ \hline
\end{tabular}
\end{center}
\end{table}

\begin{table}[h!]
\label{SI:heff_exact}
\begin{center}
\caption{TPSCI Effective Hamiltonian to show the increases singlet coupling strengths between $\ket{\mathrm{S}_1}$ and $\ket{^1\mathrm{TT}}$ in meV once we include correlations with our external space.}
\begin{tabular}{|c|c|c|c|c|c|c|c|c|c|c|c|}
\hline 
State	&	S$_0$	&	S$_1$	&	S$_1$	&	S$_1$	&	S$_1$	&	1,2	&	1,4	&	2,3	&	1,3	&	2,4	&	3,4	\\ \hline
S$_0$	&	0.00	&	-3.32	&	-0.32	&	-3.02	&	-0.52	&	-10.45	&	5.65	&	4.03	&	-1.70	&	-0.06	&	-0.03	\\
S$_1$	&	-3.32	&	0.00	&	27.38	&	-33.59	&	-14.64	&	-6.43	&	19.46	&	0.33	&	0.80	&	1.31	&	0.01	\\
S$_1$	&	-0.32	&	27.38	&	0.00	&	9.88	&	35.63	&	3.66	&	2.47	&	10.16	&	1.04	&	0.50	&	0.00	\\
S$_1$	&	-3.02	&	-33.59	&	9.88	&	0.00	&	-7.57	&	0.37	&	-0.82	&	14.36	&	-0.64	&	-0.08	&	-0.07	\\
S$_1$	&	-0.52	&	-14.64	&	35.63	&	-7.57	&	0.00	&	-0.23	&	-6.62	&	0.42	&	0.10	&	0.37	&	0.08	\\
1,2	&	-10.45	&	-6.43	&	3.66	&	0.37	&	-0.23	&	0.00	&	1.45	&	3.37	&	-2.49	&	-4.48	&	0.01	\\
1,4	&	5.65	&	19.46	&	2.47	&	-0.82	&	-6.62	&	1.45	&	0.00	&	-0.06	&	-0.05	&	11.99	&	1.47	\\
2,3	&	4.03	&	0.33	&	10.16	&	14.36	&	0.42	&	3.37	&	-0.06	&	0.00	&	10.81	&	-0.06	&	0.12	\\
1,3	&	-1.70	&	0.80	&	1.04	&	-0.64	&	0.10	&	-2.49	&	-0.05	&	10.81	&	0.00	&	-0.04	&	-5.41	\\
2,4	&	-0.06	&	1.31	&	0.50	&	-0.08	&	0.37	&	-4.48	&	11.99	&	-0.06	&	-0.04	&	0.00	&	-4.28	\\
3,4	&	-0.03	&	0.01	&	0.00	&	-0.07	&	0.08	&	0.01	&	1.47	&	0.12	&	-5.41	&	-4.28	&	0.00	\\ \hline
\end{tabular}
\end{center}
\end{table}

\section{Hexabenzocoronene}

\subsubsection{Effect of nbody and $\delta_{elec}$ on correlation energy}

The main aim of studying hexabenzocoronene is to see how inter-cluster interaction contributes to the correlation energy with the number of clusters (nbody) interacting with each other. We have done a series of TPSCI calculations by constructing the local many-body basis for each sector of Fock space which
contained up to $N_I  \pm \delta_{elec}$ number of electrons, where we varied $\delta_{elec}$ from 1 to 3 for each nbody value (2, 3, 4). For nbody=0, there is no inter-cluster interaction present in the system which is exactly cMF basis. In the nbody=1 case, only excitations inside one cluster are allowed at a time. As discussed in cMF theory section, the interaction between singly excited cluster configurations and ground state cluster configuration should be zero due to the Brillouin condition.
\begin{figure}
    \centering
    \includegraphics[width=.4\linewidth]{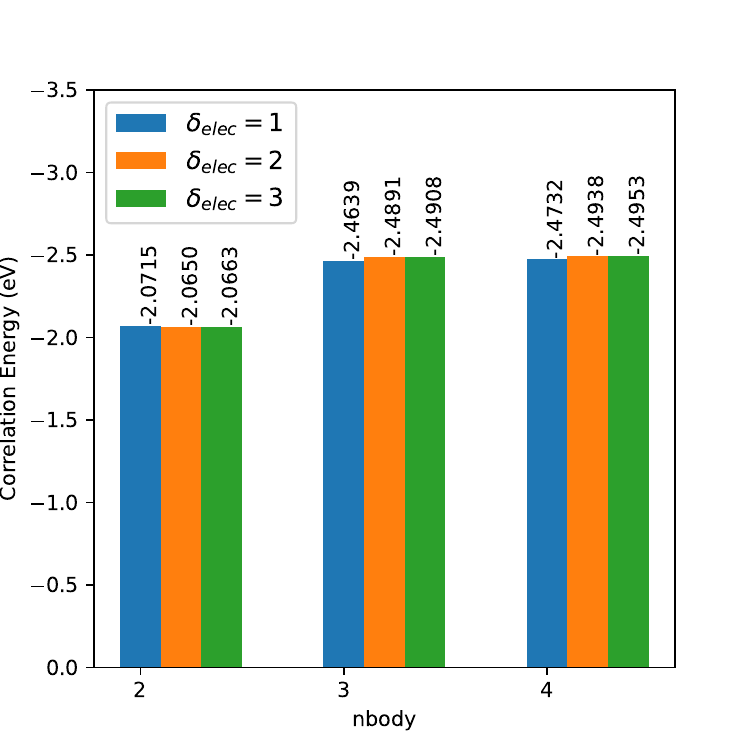}
    \caption{Effect of nbody and $\delta_{elec}$ in recovering correlation energy for hexabenzocoronene. $\epsilon_{cipsi}=2.5e{-4}$ and $\epsilon_{foi}=1e{-6}$, threshold values are used for TPSCI variational energies in STO-3g basis.}
    \label{fig:nbody_hbc}
\end{figure}

From Fig. \eqref{fig:nbody_hbc}, we can see how nbody interactions affect the correlation energy. Most of the correlation energy is recovered from 2-body interactions.
If we include 3-body interactions, it can be seen that there is 0.39-0.43 $eV$ contribution from 3-cluster interactions, and 4-body interactions further add only 0.02 $eV$ for $\delta_{elec}=2$ and nearly 0.002 $eV$ for $\delta_{elec}=3$ which is smaller than chemical accuracy.
From these observations, it can be predicted that the hexabenzocoronene system contains interactions mostly between two benzene rings although the electron density is delocalized all over the systems. However, a significant amount of contributions is achieved through the 3-body interactions.
We guess that most of the 3-body interactions are available among two benzene rings present on the outer ring and the central fused benzene ring distributed in a line (para-like) or three vertices of a triangle (ortho-like).
This can be confirmed via the computation of $3^{rd}$ order cumulant of particle number. Whereas 3 benzene rings in the outer ring along with the central benzene contribute negligibly to 4-body correlation energy. 

Now, if we focus on $\delta_{elec}$ parameter, there is a very small change (less than chemical accuracy) if we shift $\delta_{elec}$ value from 1 to 2, and 3.
$\delta_{elec}$ parameter restricts the Fock sectors ($n_{\alpha}, n_{\beta} $) based on the allowable electron transitions ($\delta_{elec}= \pm \delta$ electron transfer).
From this observation, it can be concluded that configurations created through single electron transition inside each cluster interact mostly with singly excited configurations and ground state configuration from the rest of the systems.
Inter-cluster parameter, nbody determines how many inter-cluster interactions are significant for intra-cluster parameter, $\delta_{elec}$. As $\delta_{elec}=1$ recovers most of the correlation energy for a particular nbody value, the most important physical process that happens in hexabenzocoronene must be electron hopping (charge-transfer) which is further supported by the correlation matrix of particle number operator, $\hat{N}$ and spin projection, $\hat{S}^z$.

\subsubsection{Covariance matrix and Average value of $\hat{S}^z$, $\hat{N}$, $\hat{S}^2$, $\hat{Q}$ }
Root $= 1$, Cov($N_I, N_J$):
\[
\begin{bmatrix}
0.02550829&	-0.00759149	&-0.00017893&	-0.0004038&	-0.00018108&	-0.00759755&	-0.00955544\\
    -0.00759149	&0.02551065	&-0.00759694&	-0.00018174	&-0.00040329&	-0.0001787&	-0.0095585\\
    -0.00017893	&-0.00759694&	0.02548365&	-0.00759733	&-0.00017889&	-0.00040443	&-0.00952713\\
    -0.0004038&	-0.00018174&	-0.00759733&	0.02550731	&-0.00759024&	-0.00017861	&-0.00955559\\
    -0.00018108	&-0.00040329&	-0.00017889&	-0.00759024	&0.02551021&	-0.0075972	&-0.0095595\\
    -0.00759755&	-0.0001787&	-0.00040443	&-0.00017861&	-0.0075972&	0.02548714&	-0.00953065\\
    -0.00955544	&-0.0095585&	-0.00952713	&-0.00955559&	-0.0095595&	-0.00953065	&0.05728681\\
\end{bmatrix}
\]

Root $= 1$, Cov($S^z_I, S^z_J$):
\[
\begin{bmatrix}
0.00925446 & -0.0027587 & -0.00004766 & -0.00013415 & -0.00004725 & -0.00276328 & -0.00350342 \\
-0.0027587 & 0.00925935 & -0.00276279 & -0.00004727 & -0.00013475 & -0.00004721 & -0.00350864 \\
-0.00013415 & -0.00004727 & 0.00925467 & -0.00276338 & -0.00004716 & -0.00013419 & -0.0034995 \\
-0.00004766 & -0.00276279 & -0.00276338 & 0.00925471 & -0.00275859 & -0.00004761 & -0.00350371 \\
-0.00004725 & -0.00013475 & -0.00004716 & -0.00275859 & 0.00926062 & -0.002763 & -0.00350988 \\
-0.00276328 & -0.00004721 & -0.00013419 & -0.00004761 & -0.002763 & 0.00925669 & -0.00350141 \\
-0.00350342 & -0.00350864 & -0.0034995 & -0.00350371 & -0.00350988 & -0.00350141 & 0.02102656 \\
\end{bmatrix}
\]

Root $= 1$, Cov($Q_I, Q_J$):
\[
\begin{bmatrix}
0.03307346 & 0.01079417 & 0.0001041 & -0.00005672 & 0.00011049 & 0.01080873 & 0.01342415 \\
0.01079417 & 0.03307551 & 0.01080557 & 0.00011064 & -0.00005735 & 0.00010328 & 0.01342571 \\
0.0001041 & 0.01080557 & 0.03307124 & 0.01080811 & 0.00010314 & -0.00006024 & 0.01342283 \\
-0.00005672 & 0.00011064 & 0.01080811 & 0.03307314 & 0.01079245 & 0.00010397 & 0.01342431 \\
0.00011049 & -0.00005735 & 0.00010314 & 0.01079245 & 0.03307654 & 0.01080523 & 0.01342799 \\
0.01080873 & 0.00010328 & -0.00006024 & 0.00010397 & 0.01080523 & 0.03307683 & 0.01342283 \\
0.01342415 & 0.01342571 & 0.01342283 & 0.01342431 & 0.01342799 & 0.01342283 & 0.06866612 \\
\end{bmatrix}
\]

Root $= 1$, Cov($S^2_i, S^2_j$):
\[
\begin{bmatrix}
0.03393927&	0.01005385	&-0.00014237&	-0.00009454	&-0.00013552&	0.01007677&	0.01286263\\
0.01005385	&0.03394938	&0.01007536	&-0.00013532	&-0.00009262	&-0.00014198&	0.01287511\\
-0.00014237	&0.01007536	&0.03395539&	0.01007542	&-0.00014202	&-0.00009892	&0.0128601\\
-0.00009454	&-0.00013532	&0.01007542&	0.03394001&	0.01005458	&-0.00014262&	0.012863\\
-0.00013552	&-0.00009262&	-0.00014202&	0.01005458&	0.03395483	&0.01007666&	0.0128772\\
0.01007677	&-0.00014198&	-0.00009892	&-0.00014262&	0.01007666	&0.0339672	&0.01287065\\
0.01286263	&0.01287511&	0.0128601	&0.012863&	0.0128772	&0.01287065	&0.07845006\\
\end{bmatrix}
\]

\begin{table*}[!ht]
    \caption{Average value of particle number, $\hat{N}$, spin projection, $\hat{S}^z$, $Q$, $\hat{S}^2$ for each cluster in cc-pVDZ basis.}
    \centering
    \begin{tabular}{|l|l|l|l|l|l|l|l|}
    \hline
        Cluster & 1 & 2 & 3 & 4 & 5 & 6 & 7  \\ \hline\hline
        $\langle \hat{N}\rangle$ & 6.00009508	&6.00006097	&6.00008872&	6.00009172&	6.00006299	&6.00008641	&5.9995141\\ \hline
        $\langle \hat{S}^z\rangle$& 0.00001386	&-0.0000144&	0.00000359	&0.0000157	&-0.00001414&	0.00000181	&-0.00000642 \\ \hline
        $\langle Q \rangle$ & 0.03424627&	0.03424847	&0.03424389&	0.03424593&	0.03424957&	0.03424989&	0.07416684 \\ \hline
        $\langle \hat{S}^2 \rangle$ & 0.02767142 &	0.02767881	&0.02766855	&0.02767127	& 0.02768138&	0.02767573&	0.06288520\\ \hline
    \end{tabular}
    \label{fig:table_hbc}
\end{table*}
\begin{figure}
    \centering
    \includegraphics[width=1\linewidth]{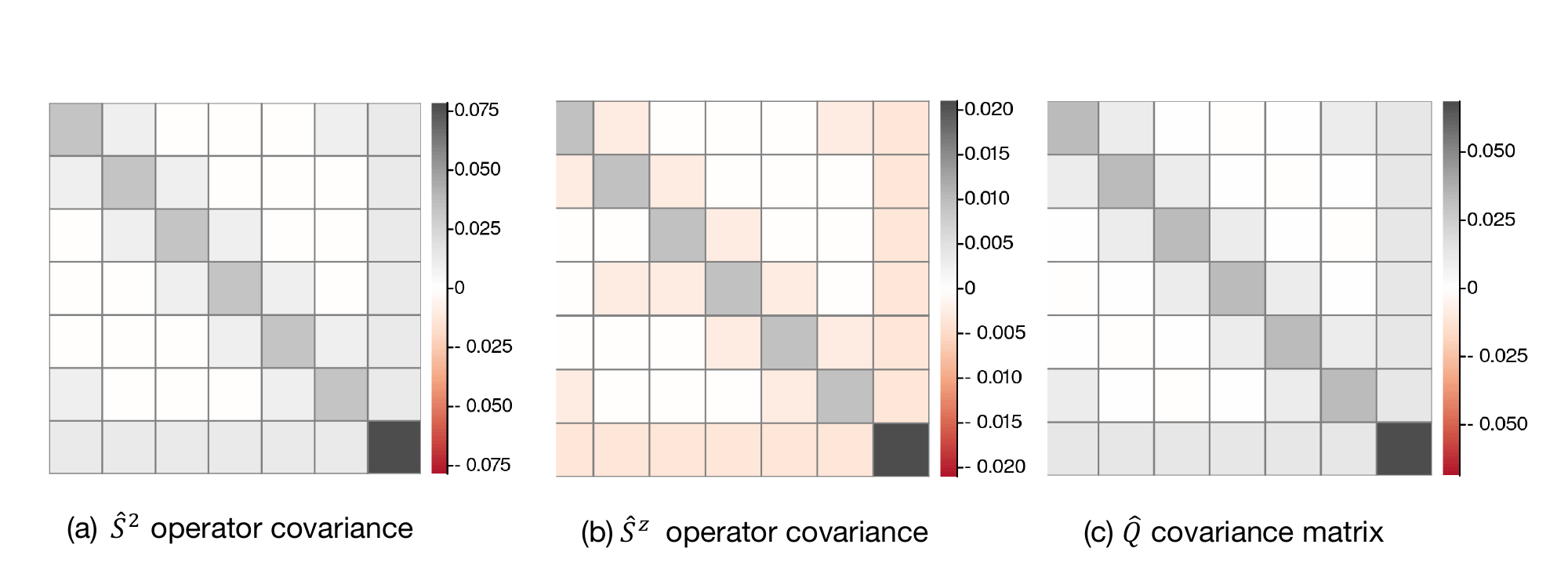}
    \caption{$\hat{S}^2$, spin projection, $\hat{S}^z$, $\hat{Q}$ covariance matrices for each cluster.}
    \label{fig:cov_mat}
\end{figure}

\subsection{Cr$_2$ orbitals}
\begin{figure}
    \centering
    \includegraphics[width=1\linewidth]{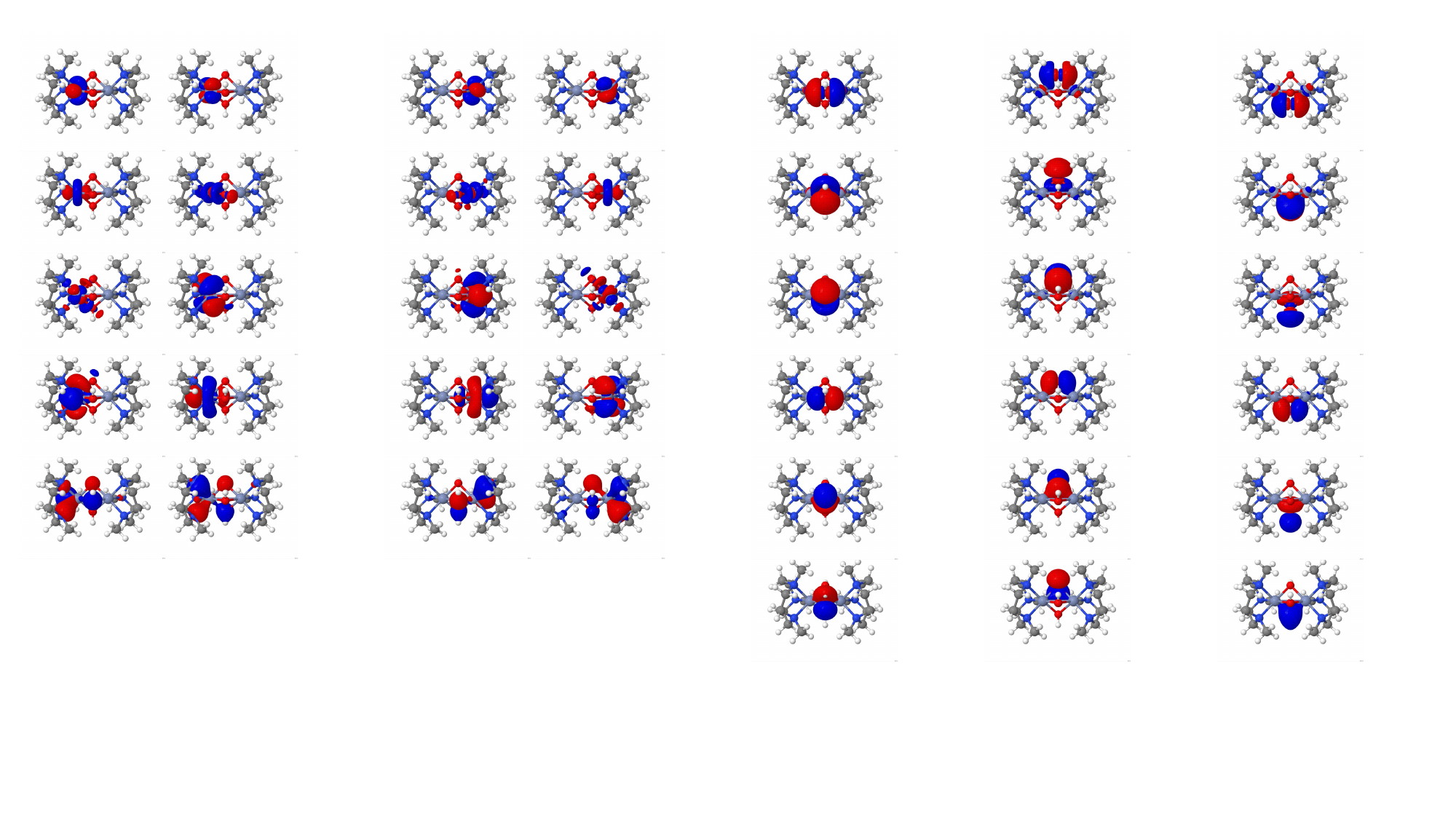}
    \caption{Molecular orbitals comprising the active site for the dichromium compound. Orbitals are grouped by cluster. }
    \label{fig:enter-label}
\end{figure}